\documentclass[lettersize,journal]{IEEEtran}

\usepackage{amsmath,amsfonts,amssymb,bm}
\usepackage{algorithmic}    
\usepackage[lined,boxed,vlined,ruled,linesnumbered]{algorithm2e}

\usepackage{graphicx}
\usepackage[caption=false,font=normalsize,labelfont=sf,textfont=sf]{subfig}
\usepackage{array}
\usepackage{booktabs}
\usepackage{tabularx}
\usepackage{adjustbox}
\usepackage{multirow}
\usepackage{colortbl}
\usepackage[table]{xcolor}
\usepackage{longtable}
\usepackage{float}

\usepackage{textcomp}
\usepackage{stfloats}
\usepackage{url}
\usepackage{verbatim}
\usepackage{hyperref}
\usepackage{balance}
\usepackage[utf8]{inputenc}
\usepackage{xspace}
\usepackage{soul, color}
\usepackage{lipsum}
\usepackage{framed}
\usepackage{listings,upquote}
\usepackage{afterpage}
\usepackage{ragged2e}

\usepackage{bbding}
\usepackage{dsfont}
\usepackage[mathscr]{eucal}
\usepackage{fontawesome5}

\usepackage{CJKutf8}
\usepackage{marginnote}
\usepackage{makecell}
\usepackage{pifont}
\usepackage{multicol}

\usepackage{forest}
\usetikzlibrary{arrows.meta,shapes,positioning,shadows,trees}
\usepackage[color,matrix,arrow,all]{xy}

\usepackage[most]{tcolorbox}
\usepackage{newlfont}
\colorlet{shadecolor}{gray!20}

\def\BibTeX{{\rm B\kern-.05em{\sc i\kern-.025em b}\kern-.08em
    T\kern-.1667em\lower.7ex\hbox{E}\kern-.125emX}}

\newcommand{\zxh}[1]{\textcolor{blue}{#1}}
\newcommand{\zw}[1]{\textcolor{purple}{zw: #1}}

\newcommand{\zj}[1]{\textcolor{magenta}{zj: #1}}
\newcommand{\hsk}[1]{\textcolor{orange}{hsk: #1}}
\newcommand{\hqk}[1]{\textcolor{blue}{hqk: #1}}
\newcommand{\why}[1]{\textcolor{cyan}{why: #1}}
\newcommand{\lzh}[1]{\textcolor{teal}{lzh: #1}}

\newcommand{\updated}{\textcolor{red}{updated}\xspace}
\newcommand{\llm}{\textsc{LLM}\xspace}
\newcommand{\llms}{\textsc{LLMs}\xspace}

\newcommand{\hi}[1]{\vspace{.25em} \noindent {\bf #1}\xspace}
\newcommand{\bfit}[1]{\textbf{\textit{#1}}}
\newcommand{\ds}{\textsc{Data Preparation}\xspace}

\newtcolorbox{takeawaybox}[1][]{
    enhanced,
    breakable,
    colback=gray!5!white,
    colframe=black!60!white,
    fonttitle=\bfseries,
    arc=1pt,
    boxrule=1pt,
    top=3pt,
    left=3pt,
    right=3pt,
    bottom=3pt,
    before skip=8pt,
    after skip=8pt,
    #1
}

\begin{document}
\bstctlcite{IEEEexample:BSTcontrol}

\title{Can LLMs Clean Up Your Mess? A Survey of Application-Ready Data Preparation with LLMs}

\author{Wei Zhou, Jun Zhou, Haoyu Wang, Zhenghao Li, Qikang He, Shaokun Han, Guoliang Li~\IEEEmembership{Fellow, IEEE}, \\  Xuanhe Zhou, Yeye He,  Chunwei Liu, Zirui Tang, Bin Wang, Shen Tang, Kai Zuo, Yuyu Luo, \\
 Zhenzhe Zheng, Conghui He, Jingren Zhou~\IEEEmembership{Fellow, IEEE}, Fan Wu \\
\hspace{0.5cm}\faGithub \hspace{0.0cm}\textit{
    {\textbf{\small Awesome-Data-LLM:}} \textcolor{blue}{\small \underline{\url{https://github.com/weAIDB/awesome-data-llm}}}
}
\thanks{Wei Zhou, Jun Zhou, Haoyu Wang, Zhenghao Li, Qikang He, Shaokun Han, Xuanhe Zhou, Zhenzhe Zheng, and Fan Wu are with Shanghai Jiao Tong University, Shanghai, China. Guoliang Li is with Tsinghua University, Beijing, China. Yeye He is with Microsoft Research. Chunwei Liu is with MIT CSAIL, USA. Bin Wang and Conghui He are with Shanghai AI Laboratory. Shen Tang and Kai Zuo are with Xiaohongshu Inc. Yuyu Luo is with the Hong Kong University of Science and Technology (Guangzhou), China. Jingren Zhou is with Alibaba Group.
}
\thanks{Corresponding author: Xuanhe Zhou (zhouxuanhe@sjtu.edu.cn).}
}

\markboth{IEEE Transactions on Knowledge and Data Engineering,~Vol.~0, No.~0, January~2025}%
{Zhou \MakeLowercase{\textit{et al.}}: Can LLMs Clean Up Your Mess? A Survey of Application-Ready Data Preparation with LLMs}

\maketitle

\begin{abstract}
Data preparation aims to denoise raw datasets, uncover cross-dataset relationships, and extract valuable insights from them, which is essential for a wide range of data-centric applications. Driven by (i) rising demands for application-ready data (e.g., for analytics, visualization, decision-making), (ii) increasingly powerful LLM techniques, and (iii) the emergence of infrastructures that facilitate flexible agent construction (e.g., using Databricks Unity Catalog), LLM-enhanced methods are rapidly becoming a transformative and potentially dominant paradigm for data preparation.


By investigating hundreds of recent literature works, this paper presents a systematic review of this evolving landscape, focusing on the use of LLM techniques to prepare data for diverse downstream tasks. First, we characterize the fundamental paradigm shift, from rule-based, model-specific pipelines to prompt-driven, context-aware, and agentic preparation workflows. Next, we introduce a task-centric taxonomy that organizes the field into three major tasks: data cleaning (e.g., standardization, error processing, imputation), data integration (e.g., entity matching, schema matching), and data enrichment (e.g., data annotation, profiling). For each task, we survey representative techniques, and highlight their respective strengths (e.g., improved generalization, semantic understanding) and limitations (e.g., the prohibitive cost of scaling LLMs, persistent hallucinations even in advanced agents, the mismatch between advanced methods and weak evaluation). Moreover, we analyze commonly used datasets and evaluation metrics (the empirical part). Finally, we discuss open research challenges and outline a forward-looking roadmap that emphasizes scalable LLM-data systems, principled designs for reliable agentic workflows, and robust evaluation protocols.
\end{abstract}


\section{Introduction}
\label{sec:introducion}

\begin{figure}[!t]
  \centering
  \includegraphics[width=\linewidth]{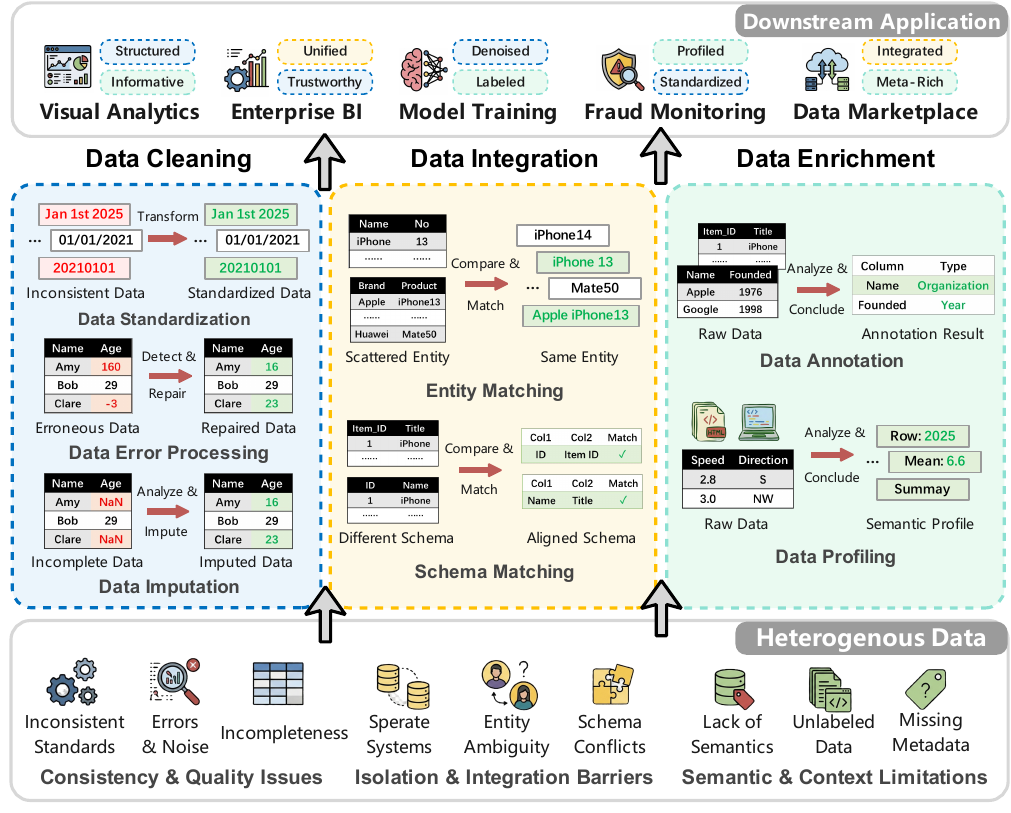}
  \caption{Application-Ready \ds\ -- \textnormal{Three core tasks (i.e., Data Cleaning, Integration, and Enrichment) address key sources of data inefficiency: quality issues, integration barriers, and semantic gaps.}
  }
  \label{fig:motivation}
  \vspace{-.65cm}
\end{figure}

\IEEEPARstart{D}{ata} preparation refers to the process of transforming raw datasets into high-quality ones (e.g., trustworthy and comprehensive) by denoising corrupted inputs, identifying cross-dataset relationships, and extracting meaningful insights. Despite its foundational role in downstream applications such as business intelligence (BI) analytics~\cite{neeli2021ensuring, Ali_Darmawan_2023}, machine learning (ML) model training~\cite{datagov4ml, DBLP:journals/corr/abs-2412-03824}, and data sharing~\cite{Guan2026, 10.1093/jamiaopen/ooaf041}, data preparation remains a critical bottleneck in real scenarios. For instance, an estimated 20\%$-$30\% of enterprise revenue is lost due to data inefficiencies~\cite{acceldata}. 
As illustrated in Figure~\ref{fig:motivation}, real-world data inefficiencies primarily arise from three sources: (1) \textit{Consistency} \& \textit{Quality Issues} (e.g., non-standard formats, noise, and incompleteness); (2) \textit{Isolation} \& \textit{Integration Barriers} (e.g., disparate systems, entity ambiguity, and schema conflicts); and (3) \textit{Semantic} \& \textit{Context Limitations} (e.g., missing metadata and unlabeled data). To these challenges, data preparation~\cite{llmdata, dataagent} involves three main tasks: \textit{Data Cleaning}, \textit{Data Integration}, and \textit{Data Enrichment}, which transform raw inputs into unified, reliable, and enriched datasets. As the volume and heterogeneity of data continue to surge (e.g., global data volume is forecast to triple from 2025 to 2029~\cite{statista}), the imperative for effective data preparation has never been greater. However, traditional data preparation methods rely heavily on static rules~\cite{LLMGDO, OpenRefine}, manual interventions, or narrowly scoped models~\cite{Ditto, TURL}, motivating the need for more intelligent, adaptive solutions.

\begin{figure*}[!t]
  \centering
  \includegraphics[width=.95\linewidth]{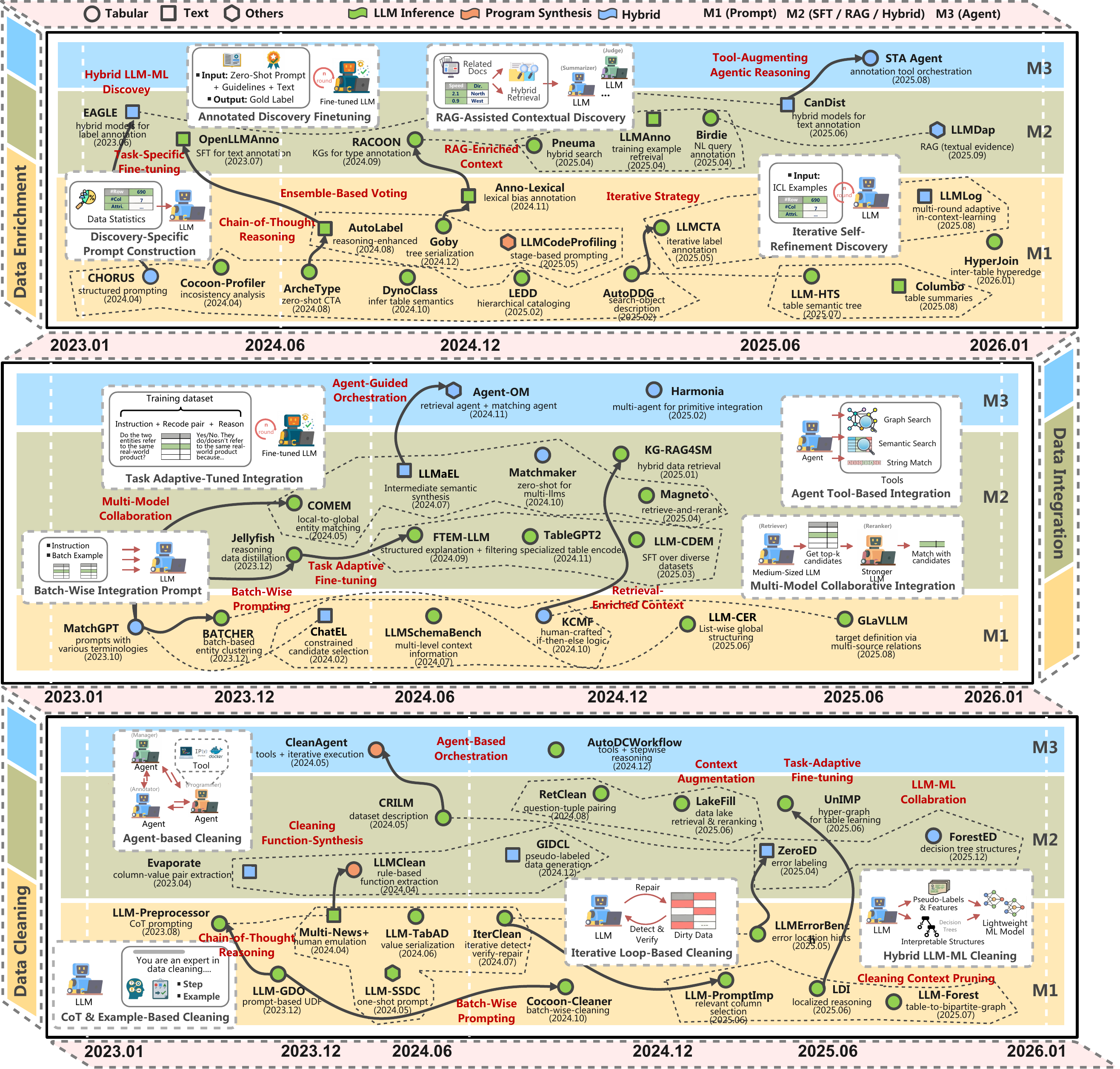}
  \caption{Overview of Application-Ready \ds through \llm-Enhanced Methods.}
  \label{fig:overview}
  \vspace{-.5cm}
\end{figure*}

\subsection{Limitations of Traditional \ds}
As discussed above, traditional preparation techniques, ranging from heuristic rule-based systems~\cite{tool2, tool1, OpenRefine} to domain-specific machine-learning models~\cite{DeepER, DeepMatcher, Ditto, TURL}, face several fundamental limitations.

\noindent $\bullet$ \textbf{(Limitation \ding{182}) High Manual Effort and Expertise Dependence.} 
Traditional data preparation methods largely depend on fixed rules and domain-specific configurations, such as regular expressions and validation constraints~\cite{AutoDCWorkflow, GIDCL}. This reliance demands substantial manual effort and specialized expert knowledge, introducing significant development and maintenance barriers. For instance, data standardization typically requires complex, handcrafted scripts (e.g., user-defined functions) or manual constraints (e.g., date formatting rules)~\cite{LLMGDO, Evaporate}. Similarly, data error processing pipelines often rely on fixed detect-then-correct workflows defined by manually crafted rules, which are not only labor-intensive to maintain but also prone to introducing new errors (e.g., incorrectly repaired values) during correction~\cite{IterClean}.


\noindent $\bullet$ \textbf{(Limitation \ding{183}) Limited Semantic Awareness in Preparation Enforcement.}
Conventional rule-based approaches predominantly rely on statistical patterns (e.g., computing missing value percentages) or syntactic matching, which fundamentally limit their ability to accurately identify complex inconsistencies that require semantic reasoning. For example, in data integration, traditional similarity-based matching techniques struggle to resolve semantic ambiguities (such as abbreviations, synonyms, or domain-specific terminology) due to the lack of commonsense or domain-specific knowledge~\cite{KGRAG4SM}. Moreover, keyword-based search mechanisms in data enrichment frequently fail to capture user intent, creating a semantic gap that leaves relevant datasets undiscovered~\cite{Birdie, Pneuma}.


\noindent $\bullet$ \textbf{(Limitation \ding{184}) Poor Generalization across Diverse Preparation Tasks and Data Modalities.} 
Traditional deep learning models typically require specialized feature engineering~\cite{TURL} or domain-specific training~\cite{Ditto}, which severely restricts their generalizability across diverse domains and data modalities. For example, fine-tuned entity-matching models exhibit significant performance degradation when applied to out-of-distribution entities~\cite{MatchGPT}. Similarly, supervised data annotation models struggle to perform well on data from underrepresented subgroups or domains with limited labeled examples~\cite{ArcheType, LLMCTA}. Furthermore, methods designed for structured tabular data often fail to effectively process semi-structured text or other modalities~\cite{SSDC}, limiting their applicability in heterogeneous data environments.


\noindent $\bullet$ \textbf{(Limitation \ding{185}) Preparation Reliance on Labeled Data and Limited Knowledge Integration.}
Small-model-based approaches typically require large volumes of high-quality and accurately labeled examples, which can be expensive to obtain at scale~\cite{Unicorn}. For instance, in data annotation, the prohibitively high cost of expert labeling limits the scale of reliable datasets, whereas crowdsourced alternatives often exhibit unstable quality~\cite{Anno-lexical}. Moreover, existing methods often lack the flexibility to integrate diverse contexts. For example, {general retrieval-based systems~\cite{Birdie} face challenges in effectively integrating structured table data with unstructured free-text context.} 



\subsection{LLM-Enhanced \ds: Driving Forces\\ And Opportunities}

To overcome these limitations, recent advances in large language models (LLMs) have catalyzed a paradigm shift in data preparation~\cite{llm4ds, empoweringtabulardatapreparation}. This transformation is fueled by three converging forces. First, the increasing demand for application-ready data, which is essential for scenarios such as personalizing customer experiences~\cite{enterprise_data_strategy} and enabling real-time analytics. Second, the methodological shift from static, rule-based pipelines to LLM agent frameworks that can autonomously plan (e.g., interpret ambiguous data patterns), execute (e.g., adapt to heterogeneous formats), and reflect on data preparation actions. Third, infrastructure advances that support flexible and cost-effective LLM technique usage, such as the API integrations for LLM agent construction in Databricks Unity Catalog~\cite{databricks} and the proliferation of open-source LLMs.


By leveraging generative capabilities, semantic reasoning, and extensive pretraining, \llms introduce a paradigm shift that offers opportunities in four aspects.



\noindent $\bullet$ \textbf{(Opportunity \ding{182}) From Manual Preparation to Instruction-Driven and Agentic Automation.}
To address the high manual effort and expertise dependence in data preparation, LLM-enhanced techniques facilitate natural-language interactions and automated workflow generation~\cite{CleanAgent, AutoDCWorkflow}. For instance, in data cleaning, users can directly define transformation logic using textual prompts rather than writing complex user-defined functions~\cite{LLMGDO}. Moreover, advanced data cleaning frameworks (e.g., Clean Agent~\cite{CleanAgent}, AutoDCWorkflow~\cite{AutoDCWorkflow}) have integrated LLM-enhanced agents to orchestrate cleaning workflows, in which agents plan and execute pipelines by identifying quality issues and invoking external tools to achieve effective data cleaning with minimal human intervention. 


\noindent $\bullet$ \textbf{(Opportunity \ding{183}) Semantic Reasoning for Consistent Preparation Enforcement.}
Unlike traditional methods that rely on syntactic similarity or heuristics, LLM-enhanced approaches incorporate semantic reasoning into preparation workflows~\cite{GIDCL, IterClean}. For example, in data integration, LLMs utilize pretrained semantic knowledge to resolve ambiguities of abbreviations, synonyms, and domain-specific terminology~\cite{KGRAG4SM}. In data enrichment, LLMs infer semantic column groups and generate human-aligned dataset descriptions, enabling more accurate dataset understanding and enrichment beyond keyword-based or statistical profiling~\cite{Cocoon, AutoDDG}.


\noindent $\bullet$ \textbf{(Opportunity \ding{184}) From Domain-Specific Preparation Training to Cross-Modal Generalization.}
LLM-enhanced techniques reduce reliance on domain-specific feature engineering and task-specific training, demonstrating strong adaptability across data modalities~\cite{Jellyfish}. For example, in data cleaning, LLMs handle heterogeneous schemas and formats by following instructions via few-shot, similarity-based in-context prompting without fine-tuning~\cite{LLMClean}. For tabular data integration, specialized encoders (e.g., TableGPT2~\cite{tablegpt2}) bridge the modality gap between tabular structures and textual queries, ensuring robust performance without extensive domain-specific feature engineering.


\noindent $\bullet$ \textbf{(Opportunity \ding{185}) Knowledge-Augmented Preparation with Minimal Labeling.} 
LLMs alleviate the need for large volumes of high-quality labels by exploiting pretrained knowledge and dynamically integrating external context~\cite{LakeFill}. For example, in entity matching, some methods incorporate external domain knowledge (e.g., from Wikidata) and structured pseudo-code into prompts to reduce reliance on task-specific training pairs~\cite{KcMF}. In data cleaning and data enrichment, Retrieval-Augmented Generation (RAG) based frameworks retrieve relevant external information from data lakes, enabling accurate value restoration and metadata generation without requiring fully observed training data~\cite{RetClean, Pneuma}.

\subsection{Contributions and Differences with Existing Surveys}


We comprehensively review recent advances in LLM-enhanced application-ready data preparation (e.g., for decision-making, analytics, or other applications) with a focused scope. Instead of covering all possible preparation tasks, we concentrate on three core tasks that appear most in existing studies~\cite{llmdata, dataagent} and real-world pipelines~\cite{realworlddp} (i.e., data cleaning, data integration, and data enrichment in Figure~\ref{fig:overview}).
Within this scope, we present a task-centered taxonomy, summarize representative methods and their technical characteristics, and discuss open problems and future research directions. 

\noindent \textbf{\underline{$\bullet$ Data Cleaning.}} 
{Targeting the \textit{Consistency} \& \textit{Quality Issues} in Figure~\ref{fig:motivation}, this task aims to produce standardized and denoised data.} We focus on three main subtasks: \emph{(1) Data Standardization}, which transforms diverse representations into unified formats using specific prompts~\cite{LLMGDO, Evaporate} or agents that automatically generate cleaning workflows~\cite{CleanAgent, AutoDCWorkflow}; \emph{(2) Data Error Processing}, which detects and repairs erroneous values (e.g., spelling mistakes, invalid values, outlier values) through direct LLM prompting~\cite{MultiNews, Cocoon, GIDCL}, methods that add context to the model~\cite{LLMClean, LLMErrorBench}, or fine-tuning models for specific error types~\cite{GIDCL}; and \emph{(3) Data Imputation}, which fills missing values using clear instructions and retrieval-augmented generation to find relevant information~\cite{RetClean}.

\noindent \textbf{\underline{$\bullet$ Data Integration.}}
{Addressing the \textit{Isolation} \& \textit{Integration Barriers} in Figure~\ref{fig:motivation}, this task aims to identify and combine related data from different sources.}
We review two core subtasks: \emph{(1) Entity Matching}, which links records referring to the same real-world entity using structured prompts~\cite{MatchGPT, BATCHER}, sometimes supported by code-based reasoning~\cite{KcMF}; and \emph{(2) Schema Matching}, which matches columns or attributes between datasets using direct prompting~\cite{LLMSchemaBench}, RAG techniques with multiple models~\cite{Magneto}, knowledge graph-based methods~\cite{KGRAG4SM}, or agent-based systems that plan the matching process~\cite{AgentOM, Harmonia}.

\noindent \textbf{\underline{$\bullet$ Data Enrichment.}}
{Focusing on the \textit{Semantic} \& \textit{Context Limitations}, this task augments datasets with semantic insights}. We cover two key subtasks: \emph{(1) Data Annotation}, which assigns data labels or types using various prompting strategies~\cite{CHORUS, Goby, LLMCTA}, supported by retrieval-based~\cite{RACOON} and LLM-generated context~\cite{Birdie}; and \emph{(2) Data Profiling}, which generates semantic profiles and summaries (e.g., metadata) using task-specific prompts~\cite{AutoDDG, LEDD}, often enhanced with external context via retrieval-augmented generation~\cite{Pneuma}.



\begin{table*}[!t]
\centering
\vspace{-.5cm}
\caption{{Technique Overview of \llm-Enhanced \ds Methods}.}
\label{tab:table-check}
\resizebox{\textwidth}{!}{
\begin{tabular}{|c|c|c|c|c|cccc|ccc|cc|c|ccc|}
\hline
\multirow{2}{*}{\textbf{Task}} &
  \multirow{2}{*}{\textbf{Category}} &
  \multirow{2}{*}{\textbf{Modality}} &
  \multirow{2}{*}{\textbf{Year}} &
  \multirow{2}{*}{\textbf{Work}} &
  \multicolumn{4}{c|}{\textbf{Prompting}} &
  \multicolumn{3}{c|}{\textbf{RAG}} &
  \multicolumn{2}{c|}{\textbf{\begin{tabular}[c]{@{}c@{}}Model\\Adaptation\end{tabular}}} &
  \multirow{2}{*}{\textbf{\begin{tabular}[c]{@{}c@{}}Agentic\\Workflow\end{tabular}}} &
  \multicolumn{3}{c|}{\textbf{Output Strategy}} \\ \cline{6-14} \cline{16-18} 
 &
   &
   &
   &
   &
  \multicolumn{1}{c|}{ICL} &
  \multicolumn{1}{c|}{CoT} &
  \multicolumn{1}{c|}{Ensemble} &
  \begin{tabular}[c]{@{}c@{}}Self-\\ Reflect\end{tabular} &
  \multicolumn{1}{c|}{Keyword} &
  \multicolumn{1}{c|}{Semantic} &
  Other &
  \multicolumn{1}{c|}{SFT} &
  RL &
   &
  \multicolumn{1}{c|}{\begin{tabular}[c]{@{}c@{}}LLM\\ Inference\end{tabular}} &
  \multicolumn{1}{c|}{\begin{tabular}[c]{@{}c@{}}Program\\ Synthesis\end{tabular}} &
  Hybrid \\ \hline
\multirow{5}{*}{\begin{tabular}[c]{@{}c@{}}Data\\ Standadization\end{tabular}} &
  \multirow{2}{*}{\begin{tabular}[c]{@{}c@{}}Prompt-Based End-to-End\\ Standardization\end{tabular}} &
  \multirow{2}{*}{Tabular} &
  2023 &
  LLMGDO~\cite{LLMGDO} &
  \multicolumn{1}{c|}{\ding{52}} &
  \multicolumn{1}{c|}{\ding{52}} &
  \multicolumn{1}{c|}{-} &
  - &
  \multicolumn{1}{c|}{-} &
  \multicolumn{1}{c|}{-} &
  - &
  \multicolumn{1}{c|}{-} &
  - &
  - &
  \multicolumn{1}{c|}{\ding{52}} &
  \multicolumn{1}{c|}{-} &
  - \\ \cline{4-18} 
 &
   &
   &
  2024 &
  LLM-Preprocessor~\cite{LLMPreprocessor} &
  \multicolumn{1}{c|}{\ding{52}} &
  \multicolumn{1}{c|}{\ding{52}} &
  \multicolumn{1}{c|}{-} &
  - &
  \multicolumn{1}{c|}{-} &
  \multicolumn{1}{c|}{-} &
  - &
  \multicolumn{1}{c|}{-} &
  - &
  - &
  \multicolumn{1}{c|}{\ding{52}} &
  \multicolumn{1}{c|}{-} &
  - \\ \cline{2-18} 
 &
  \begin{tabular}[c]{@{}c@{}}Automatic Code-Synthesis\\ Standardization\end{tabular} &
  Text &
  2023 &
  EVAPORATE~\cite{Evaporate} &
  \multicolumn{1}{c|}{\ding{52}} &
  \multicolumn{1}{c|}{-} &
  \multicolumn{1}{c|}{\ding{52}} &
  \ding{52} &
  \multicolumn{1}{c|}{\ding{52}} &
  \multicolumn{1}{c|}{-} &
  - &
  \multicolumn{1}{c|}{-} &
  - &
  - &
  \multicolumn{1}{c|}{-} &
  \multicolumn{1}{c|}{-} &
  \ding{52} \\ \cline{2-18} 
 &
  \multirow{2}{*}{\begin{tabular}[c]{@{}c@{}}Tool-Assisted Agent-Based\\ Standardization\end{tabular}} &
  \multirow{2}{*}{Tabular} &
  \multirow{2}{*}{2024} &
  AutoDCWorkflow~\cite{AutoDCWorkflow} &
  \multicolumn{1}{c|}{\ding{52}} &
  \multicolumn{1}{c|}{\ding{52}} &
  \multicolumn{1}{c|}{-} &
  \ding{52} &
  \multicolumn{1}{c|}{-} &
  \multicolumn{1}{c|}{-} &
  - &
  \multicolumn{1}{c|}{-} &
  - &
  \ding{52} &
  \multicolumn{1}{c|}{\ding{52}} &
  \multicolumn{1}{c|}{-} &
  - \\ \cline{5-18} 
 &
   &
   &
   &
  CleanAgent~\cite{CleanAgent} &
  \multicolumn{1}{c|}{\ding{52}} &
  \multicolumn{1}{c|}{\ding{52}} &
  \multicolumn{1}{c|}{-} &
  \ding{52} &
  \multicolumn{1}{c|}{-} &
  \multicolumn{1}{c|}{-} &
  - &
  \multicolumn{1}{c|}{-} &
  - &
  \ding{52} &
  \multicolumn{1}{c|}{-} &
  \multicolumn{1}{c|}{\ding{52}} &
  - \\ \hline
\multirow{10}{*}{\begin{tabular}[c]{@{}c@{}}Data Error\\ Processing\end{tabular}} &
  \multirow{5}{*}{\begin{tabular}[c]{@{}c@{}}Prompt-Based\\ End-to-End \\ Error Processing\end{tabular}} &
  \multirow{3}{*}{Tabular} &
  \multirow{2}{*}{2024} &
  Cocoon-Cleaner~\cite{Cocoon} &
  \multicolumn{1}{c|}{\ding{52}} &
  \multicolumn{1}{c|}{\ding{52}} &
  \multicolumn{1}{c|}{-} &
  - &
  \multicolumn{1}{c|}{-} &
  \multicolumn{1}{c|}{-} &
  - &
  \multicolumn{1}{c|}{-} &
  - &
  - &
  \multicolumn{1}{c|}{\ding{52}} &
  \multicolumn{1}{c|}{-} &
  - \\ \cline{5-18} 
 &
   &
   &
   &
  IterClean~\cite{IterClean} &
  \multicolumn{1}{c|}{\ding{52}} &
  \multicolumn{1}{c|}{-} &
  \multicolumn{1}{c|}{-} &
  \ding{52} &
  \multicolumn{1}{c|}{-} &
  \multicolumn{1}{c|}{-} &
  - &
  \multicolumn{1}{c|}{-} &
  - &
  - &
  \multicolumn{1}{c|}{\ding{52}} &
  \multicolumn{1}{c|}{-} &
  - \\ \cline{4-18} 
 &
   &
   &
  2025 &
  LLMErrorBench~\cite{LLMErrorBench} &
  \multicolumn{1}{c|}{\ding{52}} &
  \multicolumn{1}{c|}{-} &
  \multicolumn{1}{c|}{-} &
  \ding{52} &
  \multicolumn{1}{c|}{-} &
  \multicolumn{1}{c|}{-} &
  - &
  \multicolumn{1}{c|}{-} &
  - &
  - &
  \multicolumn{1}{c|}{\ding{52}} &
  \multicolumn{1}{c|}{-} &
  - \\ \cline{3-18} 
 &
   &
  Text &
  2024 &
  Multi-News+~\cite{MultiNews} &
  \multicolumn{1}{c|}{\ding{52}} &
  \multicolumn{1}{c|}{\ding{52}} &
  \multicolumn{1}{c|}{\ding{52}} &
  - &
  \multicolumn{1}{c|}{-} &
  \multicolumn{1}{c|}{-} &
  - &
  \multicolumn{1}{c|}{-} &
  - &
  - &
  \multicolumn{1}{c|}{\ding{52}} &
  \multicolumn{1}{c|}{-} &
  - \\ \cline{3-18} 
 &
   &
  Other &
  2024 &
  LLM-SSDC~\cite{SSDC} &
  \multicolumn{1}{c|}{\ding{52}} &
  \multicolumn{1}{c|}{-} &
  \multicolumn{1}{c|}{-} &
  - &
  \multicolumn{1}{c|}{-} &
  \multicolumn{1}{c|}{-} &
  - &
  \multicolumn{1}{c|}{-} &
  - &
  - &
  \multicolumn{1}{c|}{\ding{52}} &
  \multicolumn{1}{c|}{-} &
  - \\ \cline{2-18} 
 &
  \begin{tabular}[c]{@{}c@{}}Function-Synthesis-Oriented\\ Error Processing\end{tabular} &
  Tabular &
  2024 &
  LLMClean~\cite{LLMClean} &
  \multicolumn{1}{c|}{\ding{52}} &
  \multicolumn{1}{c|}{-} &
  \multicolumn{1}{c|}{\ding{52}} &
  - &
  \multicolumn{1}{c|}{\ding{52}} &
  \multicolumn{1}{c|}{-} &
  - &
  \multicolumn{1}{c|}{-} &
  - &
  - &
  \multicolumn{1}{c|}{-} &
  \multicolumn{1}{c|}{-} &
  \ding{52} \\ \cline{2-18} 
 &
  \multirow{2}{*}{\begin{tabular}[c]{@{}c@{}}Task Adaptive Fine-Tuned \\ Model-Based Error Processing\end{tabular}} &
  \multirow{2}{*}{Tabular} &
  \multirow{2}{*}{2024} &
  LLM-TabAD~\cite{LLM-TabAD} &
  \multicolumn{1}{c|}{\ding{52}} &
  \multicolumn{1}{c|}{-} &
  \multicolumn{1}{c|}{-} &
  - &
  \multicolumn{1}{c|}{-} &
  \multicolumn{1}{c|}{-} &
  - &
  \multicolumn{1}{c|}{\ding{52}} &
  - &
  - &
  \multicolumn{1}{c|}{\ding{52}} &
  \multicolumn{1}{c|}{-} &
  - \\ \cline{5-18} 
 &
   &
   &
   &
  GIDCL~\cite{GIDCL} &
  \multicolumn{1}{c|}{\ding{52}} &
  \multicolumn{1}{c|}{-} &
  \multicolumn{1}{c|}{-} &
  \ding{52} &
  \multicolumn{1}{c|}{-} &
  \multicolumn{1}{c|}{\ding{52}} &
  - &
  \multicolumn{1}{c|}{\ding{52}} &
  - &
  \ding{52} &
  \multicolumn{1}{c|}{-} &
  \multicolumn{1}{c|}{-} &
  \ding{52} \\ \cline{2-18} 
 &
  \multirow{2}{*}{\begin{tabular}[c]{@{}c@{}}Hybrid \llm-ML Enhanced\\ Error Processing\end{tabular}} &
  \multirow{2}{*}{Tabular} &
  \multirow{2}{*}{2025} &
  ForestED~\cite{ForestED} &
  \multicolumn{1}{c|}{\ding{52}} &
  \multicolumn{1}{c|}{-} &
  \multicolumn{1}{c|}{\ding{52}} &
  - &
  \multicolumn{1}{c|}{-} &
  \multicolumn{1}{c|}{-} &
  - &
  \multicolumn{1}{c|}{-} &
  - &
  - &
  \multicolumn{1}{c|}{-} &
  \multicolumn{1}{c|}{-} &
  \ding{52} \\ \cline{5-18} 
 &
   &
   &
   &
  ZeroED~\cite{ZeroED} &
  \multicolumn{1}{c|}{\ding{52}} &
  \multicolumn{1}{c|}{-} &
  \multicolumn{1}{c|}{-} &
  \ding{52} &
  \multicolumn{1}{c|}{-} &
  \multicolumn{1}{c|}{-} &
  - &
  \multicolumn{1}{c|}{-} &
  - &
  - &
  \multicolumn{1}{c|}{-} &
  \multicolumn{1}{c|}{-} &
  \ding{52} \\ \hline
\multirow{9}{*}{\begin{tabular}[c]{@{}c@{}}Data\\ Imputation\end{tabular}} &
  \multirow{4}{*}{\begin{tabular}[c]{@{}c@{}}Prompt-Based\\ End-to-End Imputation\end{tabular}} &
  \multirow{4}{*}{Tabular} &
  \multirow{4}{*}{2025} &
  LDI~\cite{LDI2025} &
  \multicolumn{1}{c|}{\ding{52}} &
  \multicolumn{1}{c|}{\ding{52}} &
  \multicolumn{1}{c|}{-} &
  - &
  \multicolumn{1}{c|}{-} &
  \multicolumn{1}{c|}{-} &
  - &
  \multicolumn{1}{c|}{-} &
  - &
  - &
  \multicolumn{1}{c|}{\ding{52}} &
  \multicolumn{1}{c|}{-} &
  - \\ \cline{5-18} 
 &
   &
   &
   &
  CRILM~\cite{CRILM} &
  \multicolumn{1}{c|}{-} &
  \multicolumn{1}{c|}{-} &
  \multicolumn{1}{c|}{-} &
  - &
  \multicolumn{1}{c|}{-} &
  \multicolumn{1}{c|}{-} &
  - &
  \multicolumn{1}{c|}{\ding{52}} &
  - &
  - &
  \multicolumn{1}{c|}{\ding{52}} &
  \multicolumn{1}{c|}{-} &
  - \\ \cline{5-18} 
 &
   &
   &
   &
  LLM-PromptImp~\cite{LLM-PromptImp} &
  \multicolumn{1}{c|}{\ding{52}} &
  \multicolumn{1}{c|}{-} &
  \multicolumn{1}{c|}{-} &
  - &
  \multicolumn{1}{c|}{-} &
  \multicolumn{1}{c|}{-} &
  - &
  \multicolumn{1}{c|}{-} &
  - &
  - &
  \multicolumn{1}{c|}{\ding{52}} &
  \multicolumn{1}{c|}{-} &
  - \\ \cline{5-18} 
 &
   &
   &
   &
  LLM-Forest~\cite{LLMForest} &
  \multicolumn{1}{c|}{\ding{52}} &
  \multicolumn{1}{c|}{-} &
  \multicolumn{1}{c|}{\ding{52}} &
  - &
  \multicolumn{1}{c|}{-} &
  \multicolumn{1}{c|}{-} &
  \ding{52} &
  \multicolumn{1}{c|}{-} &
  - &
  - &
  \multicolumn{1}{c|}{\ding{52}} &
  \multicolumn{1}{c|}{-} &
  - \\ \cline{2-18} 
 &
  \multirow{2}{*}{\begin{tabular}[c]{@{}c@{}}Context-Retrieval\\ Guided Imputation\end{tabular}} &
  \multirow{2}{*}{Tabular} &
  2024 &
  RetClean~\cite{RetClean} &
  \multicolumn{1}{c|}{\ding{52}} &
  \multicolumn{1}{c|}{-} &
  \multicolumn{1}{c|}{-} &
  - &
  \multicolumn{1}{c|}{\ding{52}} &
  \multicolumn{1}{c|}{\ding{52}} &
  - &
  \multicolumn{1}{c|}{-} &
  - &
  - &
  \multicolumn{1}{c|}{\ding{52}} &
  \multicolumn{1}{c|}{-} &
  - \\ \cline{4-18} 
 &
   &
   &
  2025 &
  LakeFill~\cite{LakeFill} &
  \multicolumn{1}{c|}{\ding{52}} &
  \multicolumn{1}{c|}{\ding{52}} &
  \multicolumn{1}{c|}{-} &
  - &
  \multicolumn{1}{c|}{-} &
  \multicolumn{1}{c|}{\ding{52}} &
  \ding{52} &
  \multicolumn{1}{c|}{-} &
  - &
  - &
  \multicolumn{1}{c|}{\ding{52}} &
  \multicolumn{1}{c|}{-} &
  - \\ \cline{2-18} 
 &
  \multirow{3}{*}{\begin{tabular}[c]{@{}c@{}}Model-Optimized\\ Adaptive Imputation\end{tabular}} &
  \multirow{3}{*}{Tabular} &
  2024 &
  LLM-REC~\cite{LLMREC} &
  \multicolumn{1}{c|}{-} &
  \multicolumn{1}{c|}{-} &
  \multicolumn{1}{c|}{-} &
  - &
  \multicolumn{1}{c|}{-} &
  \multicolumn{1}{c|}{-} &
  - &
  \multicolumn{1}{c|}{\ding{52}} &
  - &
  - &
  \multicolumn{1}{c|}{\ding{52}} &
  \multicolumn{1}{c|}{-} &
  - \\ \cline{4-18} 
 &
   &
   &
  \multirow{2}{*}{2025} &
  UnIMP~\cite{highorderim} &
  \multicolumn{1}{c|}{-} &
  \multicolumn{1}{c|}{-} &
  \multicolumn{1}{c|}{-} &
  - &
  \multicolumn{1}{c|}{-} &
  \multicolumn{1}{c|}{-} &
  - &
  \multicolumn{1}{c|}{\ding{52}} &
  - &
  - &
  \multicolumn{1}{c|}{\ding{52}} &
  \multicolumn{1}{c|}{-} &
  - \\ \cline{5-18} 
 &
   &
   &
   &
  Quantum-UnIMP~\cite{QuantumUnIMP2025} &
  \multicolumn{1}{c|}{-} &
  \multicolumn{1}{c|}{-} &
  \multicolumn{1}{c|}{-} &
  - &
  \multicolumn{1}{c|}{-} &
  \multicolumn{1}{c|}{-} &
  - &
  \multicolumn{1}{c|}{\ding{52}} &
  - &
  - &
  \multicolumn{1}{c|}{\ding{52}} &
  \multicolumn{1}{c|}{-} &
  - \\ \hline
\multirow{10}{*}{\begin{tabular}[c]{@{}c@{}}Entity\\ Matching\end{tabular}} &
  \multirow{5}{*}{\begin{tabular}[c]{@{}c@{}}Prompt-Based\\ End-to-End \\ Matching\end{tabular}} &
  \multirow{4}{*}{Tabular} &
  \multirow{2}{*}{2024} &
  BATCHER~\cite{BATCHER} &
  \multicolumn{1}{c|}{\ding{52}} &
  \multicolumn{1}{c|}{-} &
  \multicolumn{1}{c|}{-} &
  - &
  \multicolumn{1}{c|}{-} &
  \multicolumn{1}{c|}{-} &
  - &
  \multicolumn{1}{c|}{-} &
  - &
  - &
  \multicolumn{1}{c|}{\ding{52}} &
  \multicolumn{1}{c|}{-} &
  - \\ \cline{5-18} 
 &
   &
   &
   &
  KCMF~\cite{KcMF} &
  \multicolumn{1}{c|}{\ding{52}} &
  \multicolumn{1}{c|}{\ding{52}} &
  \multicolumn{1}{c|}{\ding{52}} &
  - &
  \multicolumn{1}{c|}{-} &
  \multicolumn{1}{c|}{\ding{52}} &
  \ding{52} &
  \multicolumn{1}{c|}{-} &
  - &
  - &
  \multicolumn{1}{c|}{-} &
  \multicolumn{1}{c|}{-} &
  \ding{52} \\ \cline{4-18} 
 &
   &
   &
  \multirow{2}{*}{2025} &
  MatchGPT~\cite{MatchGPT} &
  \multicolumn{1}{c|}{\ding{52}} &
  \multicolumn{1}{c|}{\ding{52}} &
  \multicolumn{1}{c|}{-} &
  \ding{52} &
  \multicolumn{1}{c|}{-} &
  \multicolumn{1}{c|}{-} &
  - &
  \multicolumn{1}{c|}{\ding{52}} &
  - &
  - &
  \multicolumn{1}{c|}{-} &
  \multicolumn{1}{c|}{-} &
  \ding{52} \\ \cline{5-18} 
 &
   &
   &
   &
  LLM-CER~\cite{LLM-CER} &
  \multicolumn{1}{c|}{\ding{52}} &
  \multicolumn{1}{c|}{-} &
  \multicolumn{1}{c|}{-} &
  \ding{52} &
  \multicolumn{1}{c|}{-} &
  \multicolumn{1}{c|}{-} &
  - &
  \multicolumn{1}{c|}{-} &
  - &
  - &
  \multicolumn{1}{c|}{-} &
  \multicolumn{1}{c|}{\ding{52}} &
  - \\ \cline{3-18} 
 &
   &
  Text &
  2024 &
  ChatEL~\cite{ChatEL} &
  \multicolumn{1}{c|}{\ding{52}} &
  \multicolumn{1}{c|}{-} &
  \multicolumn{1}{c|}{-} &
  - &
  \multicolumn{1}{c|}{-} &
  \multicolumn{1}{c|}{-} &
  - &
  \multicolumn{1}{c|}{-} &
  - &
  - &
  \multicolumn{1}{c|}{\ding{52}} &
  \multicolumn{1}{c|}{-} &
  - \\ \cline{2-18} 
 &
  \multirow{3}{*}{\begin{tabular}[c]{@{}c@{}}Task-Adaptive-Tuned\\ Matching\end{tabular}} &
  \multirow{3}{*}{Tabular} &
  2023 &
  Jellyfish~\cite{Jellyfish} &
  \multicolumn{1}{c|}{-} &
  \multicolumn{1}{c|}{\ding{52}} &
  \multicolumn{1}{c|}{-} &
  - &
  \multicolumn{1}{c|}{-} &
  \multicolumn{1}{c|}{-} &
  - &
  \multicolumn{1}{c|}{\ding{52}} &
  - &
  - &
  \multicolumn{1}{c|}{\ding{52}} &
  \multicolumn{1}{c|}{-} &
  - \\ \cline{4-18} 
 &
   &
   &
  \multirow{2}{*}{2025} &
  LLM-CDEM~\cite{LLM-CDEM} &
  \multicolumn{1}{c|}{\ding{52}} &
  \multicolumn{1}{c|}{-} &
  \multicolumn{1}{c|}{-} &
  - &
  \multicolumn{1}{c|}{-} &
  \multicolumn{1}{c|}{-} &
  - &
  \multicolumn{1}{c|}{\ding{52}} &
  - &
  - &
  \multicolumn{1}{c|}{\ding{52}} &
  \multicolumn{1}{c|}{-} &
  - \\ \cline{5-18} 
 &
   &
   &
   &
  FTEM-LLM~\cite{FTEM-LLM} &
  \multicolumn{1}{c|}{-} &
  \multicolumn{1}{c|}{-} &
  \multicolumn{1}{c|}{-} &
  - &
  \multicolumn{1}{c|}{-} &
  \multicolumn{1}{c|}{-} &
  - &
  \multicolumn{1}{c|}{\ding{52}} &
  - &
  - &
  \multicolumn{1}{c|}{\ding{52}} &
  \multicolumn{1}{c|}{-} &
  - \\ \cline{2-18} 
 &
  \multirow{2}{*}{\begin{tabular}[c]{@{}c@{}}Multi-Model\\ Collaborative Matching\end{tabular}} &
  Tabular &
  2025 &
  COMEM~\cite{COMEM} &
  \multicolumn{1}{c|}{\ding{52}} &
  \multicolumn{1}{c|}{-} &
  \multicolumn{1}{c|}{-} &
  - &
  \multicolumn{1}{c|}{-} &
  \multicolumn{1}{c|}{-} &
  - &
  \multicolumn{1}{c|}{-} &
  - &
  - &
  \multicolumn{1}{c|}{\ding{52}} &
  \multicolumn{1}{c|}{-} &
  - \\ \cline{3-18} 
 &
   &
  Text &
  2025 &
  LLMaEL~\cite{LLMaEL} &
  \multicolumn{1}{c|}{\ding{52}} &
  \multicolumn{1}{c|}{-} &
  \multicolumn{1}{c|}{\ding{52}} &
  - &
  \multicolumn{1}{c|}{-} &
  \multicolumn{1}{c|}{-} &
  - &
  \multicolumn{1}{c|}{-} &
  - &
  - &
  \multicolumn{1}{c|}{-} &
  \multicolumn{1}{c|}{-} &
  \ding{52} \\ \hline
\multirow{9}{*}{\begin{tabular}[c]{@{}c@{}}Schema\\ Matching\end{tabular}} &
  \multirow{2}{*}{\begin{tabular}[c]{@{}c@{}}Prompt-Based \\ End-to-End Matching\end{tabular}} &
  \multirow{2}{*}{Tabular} &
  2024 &
  LLMSchemaBench~\cite{LLMSchemaBench} &
  \multicolumn{1}{c|}{-} &
  \multicolumn{1}{c|}{\ding{52}} &
  \multicolumn{1}{c|}{-} &
  - &
  \multicolumn{1}{c|}{-} &
  \multicolumn{1}{c|}{-} &
  - &
  \multicolumn{1}{c|}{-} &
  - &
  - &
  \multicolumn{1}{c|}{\ding{52}} &
  \multicolumn{1}{c|}{-} &
  - \\ \cline{4-18} 
 &
   &
   &
  2025 &
  GLaVLLM~\cite{GLaVLLM} &
  \multicolumn{1}{c|}{-} &
  \multicolumn{1}{c|}{-} &
  \multicolumn{1}{c|}{\ding{52}} &
  - &
  \multicolumn{1}{c|}{-} &
  \multicolumn{1}{c|}{-} &
  - &
  \multicolumn{1}{c|}{-} &
  - &
  - &
  \multicolumn{1}{c|}{\ding{52}} &
  \multicolumn{1}{c|}{-} &
  - \\ \cline{2-18} 
 &
  \multirow{2}{*}{\begin{tabular}[c]{@{}c@{}}Retrieval-Enriched\\ Contextual Matching\end{tabular}} &
  \multirow{2}{*}{Tabular} &
  2024 &
  Matchmaker~\cite{Matchmaker} &
  \multicolumn{1}{c|}{\ding{52}} &
  \multicolumn{1}{c|}{-} &
  \multicolumn{1}{c|}{-} &
  \ding{52} &
  \multicolumn{1}{c|}{-} &
  \multicolumn{1}{c|}{\ding{52}} &
  - &
  \multicolumn{1}{c|}{-} &
  - &
  \ding{52} &
  \multicolumn{1}{c|}{-} &
  \multicolumn{1}{c|}{-} &
  \ding{52} \\ \cline{4-18} 
 &
   &
   &
  2025 &
  KG-RAG4SM~\cite{KGRAG4SM} &
  \multicolumn{1}{c|}{\ding{52}} &
  \multicolumn{1}{c|}{-} &
  \multicolumn{1}{c|}{\ding{52}} &
  - &
  \multicolumn{1}{c|}{-} &
  \multicolumn{1}{c|}{\ding{52}} &
  - &
  \multicolumn{1}{c|}{-} &
  - &
  - &
  \multicolumn{1}{c|}{\ding{52}} &
  \multicolumn{1}{c|}{-} &
  - \\ \cline{2-18} 
 &
  \multirow{2}{*}{\begin{tabular}[c]{@{}c@{}}Model-Optimized\\ Adaptive Matching\end{tabular}} &
  \multirow{2}{*}{Tabular} &
  \multirow{2}{*}{2024} &
  TableLlama~\cite{tablellama} &
  \multicolumn{1}{c|}{-} &
  \multicolumn{1}{c|}{-} &
  \multicolumn{1}{c|}{-} &
  - &
  \multicolumn{1}{c|}{-} &
  \multicolumn{1}{c|}{-} &
  - &
  \multicolumn{1}{c|}{\ding{52}} &
  - &
  - &
  \multicolumn{1}{c|}{\ding{52}} &
  \multicolumn{1}{c|}{-} &
  - \\ \cline{5-18} 
 &
   &
   &
   &
  TableGPT2~\cite{tablegpt2} &
  \multicolumn{1}{c|}{\ding{52}} &
  \multicolumn{1}{c|}{\ding{52}} &
  \multicolumn{1}{c|}{-} &
  \ding{52} &
  \multicolumn{1}{c|}{-} &
  \multicolumn{1}{c|}{\ding{52}} &
  \ding{52} &
  \multicolumn{1}{c|}{\ding{52}} &
  - &
  \ding{52} &
  \multicolumn{1}{c|}{-} &
  \multicolumn{1}{c|}{-} &
  \ding{52} \\ \cline{2-18} 
 &
  \begin{tabular}[c]{@{}c@{}}Multi-Model\\ Collaborative Matching\end{tabular} &
  Tabular &
  2025 &
  Magneto~\cite{Magneto} &
  \multicolumn{1}{c|}{-} &
  \multicolumn{1}{c|}{-} &
  \multicolumn{1}{c|}{-} &
  - &
  \multicolumn{1}{c|}{-} &
  \multicolumn{1}{c|}{\ding{52}} &
  - &
  \multicolumn{1}{c|}{\ding{52}} &
  - &
  - &
  \multicolumn{1}{c|}{\ding{52}} &
  \multicolumn{1}{c|}{-} &
  - \\ \cline{2-18} 
 &
  \multirow{2}{*}{\begin{tabular}[c]{@{}c@{}}Agent-Guided\\ Orchestration-based Matching\end{tabular}} &
  Other &
  2024 &
  Agent-OM~\cite{AgentOM} &
  \multicolumn{1}{c|}{\ding{52}} &
  \multicolumn{1}{c|}{\ding{52}} &
  \multicolumn{1}{c|}{-} &
  \ding{52} &
  \multicolumn{1}{c|}{-} &
  \multicolumn{1}{c|}{\ding{52}} &
  \ding{52} &
  \multicolumn{1}{c|}{-} &
  - &
  \ding{52} &
  \multicolumn{1}{c|}{-} &
  \multicolumn{1}{c|}{-} &
  \ding{52} \\ \cline{3-18} 
 &
   &
  Tabular &
  2025 &
  Harmonia~\cite{Harmonia} &
  \multicolumn{1}{c|}{-} &
  \multicolumn{1}{c|}{\ding{52}} &
  \multicolumn{1}{c|}{-} &
  \ding{52} &
  \multicolumn{1}{c|}{-} &
  \multicolumn{1}{c|}{-} &
  - &
  \multicolumn{1}{c|}{-} &
  - &
  \ding{52} &
  \multicolumn{1}{c|}{-} &
  \multicolumn{1}{c|}{-} &
  \ding{52} \\ \hline
\multirow{18}{*}{\begin{tabular}[c]{@{}c@{}}Data\\ Annotation\end{tabular}} &
  \multirow{9}{*}{\begin{tabular}[c]{@{}c@{}}Prompt-based\\ End-to-End\\ Annotation\end{tabular}} &
  \multirow{5}{*}{Tabular} &
  \multirow{3}{*}{2024} &
  ArcheType~\cite{ArcheType} &
  \multicolumn{1}{c|}{\ding{52}} &
  \multicolumn{1}{c|}{-} &
  \multicolumn{1}{c|}{-} &
  - &
  \multicolumn{1}{c|}{-} &
  \multicolumn{1}{c|}{-} &
  - &
  \multicolumn{1}{c|}{-} &
  - &
  - &
  \multicolumn{1}{c|}{\ding{52}} &
  \multicolumn{1}{c|}{-} &
  - \\ \cline{5-18} 
 &
   &
   &
   &
  CHORUS~\cite{CHORUS} &
  \multicolumn{1}{c|}{\ding{52}} &
  \multicolumn{1}{c|}{-} &
  \multicolumn{1}{c|}{-} &
  - &
  \multicolumn{1}{c|}{-} &
  \multicolumn{1}{c|}{-} &
  - &
  \multicolumn{1}{c|}{-} &
  - &
  - &
  \multicolumn{1}{c|}{-} &
  \multicolumn{1}{c|}{-} &
  \ding{52} \\ \cline{5-18} 
 &
   &
   &
   &
  Goby~\cite{Goby} &
  \multicolumn{1}{c|}{\ding{52}} &
  \multicolumn{1}{c|}{\ding{52}} &
  \multicolumn{1}{c|}{-} &
  - &
  \multicolumn{1}{c|}{-} &
  \multicolumn{1}{c|}{-} &
  - &
  \multicolumn{1}{c|}{-} &
  - &
  - &
  \multicolumn{1}{c|}{\ding{52}} &
  \multicolumn{1}{c|}{-} &
  - \\ \cline{4-18} 
 &
   &
   &
  \multirow{2}{*}{2025} &
  Columbo~\cite{Columbo} &
  \multicolumn{1}{c|}{\ding{52}} &
  \multicolumn{1}{c|}{\ding{52}} &
  \multicolumn{1}{c|}{-} &
  - &
  \multicolumn{1}{c|}{-} &
  \multicolumn{1}{c|}{-} &
  - &
  \multicolumn{1}{c|}{-} &
  - &
  - &
  \multicolumn{1}{c|}{\ding{52}} &
  \multicolumn{1}{c|}{-} &
  - \\ \cline{5-18} 
 &
   &
   &
   &
  LLMCTA~\cite{LLMCTA} &
  \multicolumn{1}{c|}{\ding{52}} &
  \multicolumn{1}{c|}{-} &
  \multicolumn{1}{c|}{-} &
  \ding{52} &
  \multicolumn{1}{c|}{-} &
  \multicolumn{1}{c|}{-} &
  - &
  \multicolumn{1}{c|}{\ding{52}} &
  - &
  - &
  \multicolumn{1}{c|}{\ding{52}} &
  \multicolumn{1}{c|}{-} &
  - \\ \cline{3-18} 
 &
   &
  \multirow{4}{*}{Text} &
  2023 &
  EAGLE~\cite{EAGLE} &
  \multicolumn{1}{c|}{-} &
  \multicolumn{1}{c|}{-} &
  \multicolumn{1}{c|}{-} &
  - &
  \multicolumn{1}{c|}{-} &
  \multicolumn{1}{c|}{-} &
  - &
  \multicolumn{1}{c|}{\ding{52}} &
  - &
  - &
  \multicolumn{1}{c|}{\ding{52}} &
  \multicolumn{1}{c|}{-} &
  \ding{52} \\ \cline{4-18} 
 &
   &
   &
  2024 &
  AutoLabel~\cite{AutoLabel} &
  \multicolumn{1}{c|}{\ding{52}} &
  \multicolumn{1}{c|}{\ding{52}} &
  \multicolumn{1}{c|}{-} &
  - &
  \multicolumn{1}{c|}{-} &
  \multicolumn{1}{c|}{-} &
  - &
  \multicolumn{1}{c|}{-} &
  - &
  - &
  \multicolumn{1}{c|}{\ding{52}} &
  \multicolumn{1}{c|}{-} &
  - \\ \cline{4-18} 
 &
   &
   &
  \multirow{2}{*}{2025} &
  LLMLog~\cite{LLMLog} &
  \multicolumn{1}{c|}{\ding{52}} &
  \multicolumn{1}{c|}{-} &
  \multicolumn{1}{c|}{-} &
  - &
  \multicolumn{1}{c|}{-} &
  \multicolumn{1}{c|}{-} &
  - &
  \multicolumn{1}{c|}{-} &
  - &
  - &
  \multicolumn{1}{c|}{\ding{52}} &
  \multicolumn{1}{c|}{-} &
  - \\ \cline{5-18} 
 &
   &
   &
   &
  Anno-Lexical~\cite{Anno-lexical} &
  \multicolumn{1}{c|}{\ding{52}} &
  \multicolumn{1}{c|}{\ding{52}} &
  \multicolumn{1}{c|}{\ding{52}} &
  - &
  \multicolumn{1}{c|}{\ding{52}} &
  \multicolumn{1}{c|}{-} &
  - &
  \multicolumn{1}{c|}{-} &
  - &
  - &
  \multicolumn{1}{c|}{\ding{52}} &
  \multicolumn{1}{c|}{-} &
  - \\ \cline{2-18} 
 &
  \multirow{3}{*}{\begin{tabular}[c]{@{}c@{}}RAG-Assisted\\ Contextual Annotation\end{tabular}} &
  \multirow{2}{*}{Tabular} &
  2024 &
  RACOON~\cite{RACOON} &
  \multicolumn{1}{c|}{-} &
  \multicolumn{1}{c|}{-} &
  \multicolumn{1}{c|}{-} &
  - &
  \multicolumn{1}{c|}{-} &
  \multicolumn{1}{c|}{-} &
  \ding{52} &
  \multicolumn{1}{c|}{-} &
  - &
  - &
  \multicolumn{1}{c|}{\ding{52}} &
  \multicolumn{1}{c|}{-} &
  - \\ \cline{4-18} 
 &
   &
   &
  2025 &
  Birdie~\cite{Birdie} &
  \multicolumn{1}{c|}{-} &
  \multicolumn{1}{c|}{-} &
  \multicolumn{1}{c|}{-} &
  - &
  \multicolumn{1}{c|}{\ding{52}} &
  \multicolumn{1}{c|}{-} &
  - &
  \multicolumn{1}{c|}{\ding{52}} &
  - &
  - &
  \multicolumn{1}{c|}{\ding{52}} &
  \multicolumn{1}{c|}{-} &
  - 
  \\ \cline{3-18} 
 &
   &
  Text &
  2025 &
  LLMAnno~\cite{LLMAnno} &
  \multicolumn{1}{c|}{\ding{52}} &
  \multicolumn{1}{c|}{-} &
  \multicolumn{1}{c|}{-} &
  - &
  \multicolumn{1}{c|}{-} &
  \multicolumn{1}{c|}{\ding{52}} &
  - &
  \multicolumn{1}{c|}{-} &
  - &
  - &
  \multicolumn{1}{c|}{\ding{52}} &
  \multicolumn{1}{c|}{-} &
  - \\ \cline{2-18} 
 &
  \multirow{2}{*}{\begin{tabular}[c]{@{}c@{}}Fine-tuned Augmented\\ Annotation\end{tabular}} &
  Tabular &
  2025 &
  PACTA~\cite{PACTA} &
  \multicolumn{1}{c|}{\ding{52}} &
  \multicolumn{1}{c|}{-} &
  \multicolumn{1}{c|}{-} &
  - &
  \multicolumn{1}{c|}{-} &
  \multicolumn{1}{c|}{-} &
  - &
  \multicolumn{1}{c|}{\ding{52}} &
  - &
  - &
  \multicolumn{1}{c|}{\ding{52}} &
  \multicolumn{1}{c|}{-} &
  - \\ \cline{3-18} 
 &
   &
  Text &
  2025 &
  OpenLLMAnno~\cite{OpenLLMAnno} &
  \multicolumn{1}{c|}{\ding{52}} &
  \multicolumn{1}{c|}{\ding{52}} &
  \multicolumn{1}{c|}{-} &
  - &
  \multicolumn{1}{c|}{-} &
  \multicolumn{1}{c|}{-} &
  - &
  \multicolumn{1}{c|}{\ding{52}} &
  - &
  - &
  \multicolumn{1}{c|}{\ding{52}} &
  \multicolumn{1}{c|}{-} &
  - \\ \cline{2-18} 
 &
  \multirow{2}{*}{Hybrid \llm-ML Annotation} &
  \multirow{2}{*}{Text} &
  \multirow{2}{*}{2025} &
  CanDist~\cite{CanDist} &
  \multicolumn{1}{c|}{\ding{52}} &
  \multicolumn{1}{c|}{-} &
  \multicolumn{1}{c|}{-} &
  - &
  \multicolumn{1}{c|}{-} &
  \multicolumn{1}{c|}{-} &
  - &
  \multicolumn{1}{c|}{\ding{52}} &
  - &
  - &
  \multicolumn{1}{c|}{-} &
  \multicolumn{1}{c|}{-} &
  \ding{52} \\ \cline{5-18} 
 &
   &
   &
   &
  AutoAnnotator~\cite{AutoAnnotator} &
  \multicolumn{1}{c|}{-} &
  \multicolumn{1}{c|}{-} &
  \multicolumn{1}{c|}{\ding{52}} &
  - &
  \multicolumn{1}{c|}{-} &
  \multicolumn{1}{c|}{-} &
  - &
  \multicolumn{1}{c|}{\ding{52}} &
  - &
  - &
  \multicolumn{1}{c|}{-} &
  \multicolumn{1}{c|}{-} &
  \ding{52} \\ \cline{2-18} 
 &
  \multirow{2}{*}{\begin{tabular}[c]{@{}c@{}}Tool-Assisted \\ Agent-based Annotation\end{tabular}} &
  Tabular &
  2025 &
  STA Agent~\cite{STA-Agent} &
  \multicolumn{1}{c|}{-} &
  \multicolumn{1}{c|}{-} &
  \multicolumn{1}{c|}{-} &
  - &
  \multicolumn{1}{c|}{-} &
  \multicolumn{1}{c|}{\ding{52}} &
  - &
  \multicolumn{1}{c|}{-} &
  - &
  \ding{52} &
  \multicolumn{1}{c|}{-} &
  \multicolumn{1}{c|}{-} &
  \ding{52} \\ \cline{3-18} 
 &
   &
  Other &
  2025 &
  TESSA~\cite{TESSA} &
  \multicolumn{1}{c|}{\ding{52}} &
  \multicolumn{1}{c|}{-} &
  \multicolumn{1}{c|}{-} &
  \ding{52} &
  \multicolumn{1}{c|}{-} &
  \multicolumn{1}{c|}{-} &
  - &
  \multicolumn{1}{c|}{-} &
  \ding{52} &
  \ding{52} &
  \multicolumn{1}{c|}{\ding{52}} &
  \multicolumn{1}{c|}{-} &
  - \\ \hline
\multirow{10}{*}{\begin{tabular}[c]{@{}c@{}}Data\\ Profiling\end{tabular}} &
  \multirow{8}{*}{\begin{tabular}[c]{@{}c@{}}Prompt-Based\\ End-to-End\\ Profiling\end{tabular}} &
  \multirow{7}{*}{Tabular} &
  \multirow{2}{*}{2024} &
  DynoClass~\cite{DynoClass} &
  \multicolumn{1}{c|}{\ding{52}} &
  \multicolumn{1}{c|}{-} &
  \multicolumn{1}{c|}{-} &
  - &
  \multicolumn{1}{c|}{-} &
  \multicolumn{1}{c|}{-} &
  - &
  \multicolumn{1}{c|}{-} &
  - &
  - &
  \multicolumn{1}{c|}{\ding{52}} &
  \multicolumn{1}{c|}{-} &
  - \\ \cline{5-18} 
 &
   &
   &
   &
  Cocoon~\cite{10.1145/3665939.3665957} &
  \multicolumn{1}{c|}{\ding{52}} &
  \multicolumn{1}{c|}{\ding{52}} &
  \multicolumn{1}{c|}{-} &
  - &
  \multicolumn{1}{c|}{-} &
  \multicolumn{1}{c|}{-} &
  - &
  \multicolumn{1}{c|}{-} &
  - &
  - &
  \multicolumn{1}{c|}{\ding{52}} &
  \multicolumn{1}{c|}{-} &
  - \\ \cline{4-18} 
 &
   &
   &
  \multirow{3}{*}{2025} &
  AutoDDG~\cite{AutoDDG} &
  \multicolumn{1}{c|}{\ding{52}} &
  \multicolumn{1}{c|}{-} &
  \multicolumn{1}{c|}{-} &
  - &
  \multicolumn{1}{c|}{-} &
  \multicolumn{1}{c|}{-} &
  - &
  \multicolumn{1}{c|}{-} &
  - &
  - &
  \multicolumn{1}{c|}{\ding{52}} &
  \multicolumn{1}{c|}{-} &
  - \\ \cline{5-18} 
 &
   &
   &
   &
  LEDD~\cite{LEDD} &
  \multicolumn{1}{c|}{\ding{52}} &
  \multicolumn{1}{c|}{-} &
  \multicolumn{1}{c|}{-} &
  - &
  \multicolumn{1}{c|}{-} &
  \multicolumn{1}{c|}{-} &
  - &
  \multicolumn{1}{c|}{-} &
  - &
  - &
  \multicolumn{1}{c|}{\ding{52}} &
  \multicolumn{1}{c|}{-} &
  - \\ \cline{5-18} 
 &
   &
   &
   &
  LLM-HTS~\cite{LLM-HTS} &
  \multicolumn{1}{c|}{\ding{52}} &
  \multicolumn{1}{c|}{-} &
  \multicolumn{1}{c|}{-} &
  - &
  \multicolumn{1}{c|}{-} &
  \multicolumn{1}{c|}{-} &
  - &
  \multicolumn{1}{c|}{-} &
  - &
  - &
  \multicolumn{1}{c|}{\ding{52}} &
  \multicolumn{1}{c|}{-} &
  - \\ \cline{4-18} 
 &
   &
   &
  \multirow{2}{*}{2026} &
  HyperJoin~\cite{HyperJoin} &
  \multicolumn{1}{c|}{\ding{52}} &
  \multicolumn{1}{c|}{-} &
  \multicolumn{1}{c|}{-} &
  - &
  \multicolumn{1}{c|}{-} &
  \multicolumn{1}{c|}{\ding{52}} &
  - &
  \multicolumn{1}{c|}{-} &
  - &
  - &
  \multicolumn{1}{c|}{\ding{52}} &
  \multicolumn{1}{c|}{-} &
  - \\ \cline{5-18} 
 &
   &
   &
   &
  OCTOPUS~\cite{Octopus} &
  \multicolumn{1}{c|}{\ding{52}} &
  \multicolumn{1}{c|}{-} &
  \multicolumn{1}{c|}{-} &
  \ding{52} &
  \multicolumn{1}{c|}{\ding{52}} &
  \multicolumn{1}{c|}{\ding{52}} &
  - &
  \multicolumn{1}{c|}{-} &
  - &
  - &
  \multicolumn{1}{c|}{-} &
  \multicolumn{1}{c|}{\ding{52}} &
  - \\ \cline{3-18} 
 &
   &
  Other &
  2025 &
  LLMCodeProfiling~\cite{LLMCodeProfiling} &
  \multicolumn{1}{c|}{\ding{52}} &
  \multicolumn{1}{c|}{\ding{52}} &
  \multicolumn{1}{c|}{-} &
  - &
  \multicolumn{1}{c|}{-} &
  \multicolumn{1}{c|}{-} &
  - &
  \multicolumn{1}{c|}{-} &
  - &
  - &
  \multicolumn{1}{c|}{-} &
  \multicolumn{1}{c|}{\ding{52}} &
  - \\ \cline{2-18} 
 &
  \multirow{2}{*}{\begin{tabular}[c]{@{}c@{}}RAG-Assisted\\ Contextual Profiling\end{tabular}} &
  Tabular &
  2025 &
  Pneuma~\cite{Pneuma} &
  \multicolumn{1}{c|}{\ding{52}} &
  \multicolumn{1}{c|}{-} &
  \multicolumn{1}{c|}{-} &
  - &
  \multicolumn{1}{c|}{\ding{52}} &
  \multicolumn{1}{c|}{\ding{52}} &
  - &
  \multicolumn{1}{c|}{-} &
  - &
  - &
  \multicolumn{1}{c|}{\ding{52}} &
  \multicolumn{1}{c|}{-} &
  - \\ \cline{3-18} 
 &
   &
  Other &
  2025 &
  LLMDap~\cite{LLMDap} &
  \multicolumn{1}{c|}{\ding{52}} &
  \multicolumn{1}{c|}{-} &
  \multicolumn{1}{c|}{-} &
  - &
  \multicolumn{1}{c|}{-} &
  \multicolumn{1}{c|}{\ding{52}} &
  \ding{52} &
  \multicolumn{1}{c|}{-} &
  - &
  - &
  \multicolumn{1}{c|}{\ding{52}} &
  \multicolumn{1}{c|}{-} &
  - \\ \hline
\end{tabular}

}
\vspace{-.5cm}
\end{table*}

\begingroup
\color{black}

{Compared with existing \llm and data preparation surveys{~\cite{annotationsurvey, ding2024dataaugmentationusinglarge, cheng2025survey, llm4ds, datajuicersurvey, Nad_Synthetic_2025, empoweringtabulardatapreparation, shi2025surveysynthetictabular, codesurvey, long2024llms}}, our survey differs in several significant aspects.}

\noindent $\bullet$ \textbf{Holistic \textit{vs.} Limited Task Scope.}
We provide a comprehensive review of three fundamental data preparation tasks (cleaning, integration, enrichment) across diverse data modalities, including table and text.
In contrast, existing surveys typically limit their scope to specific tasks~\cite{Heterogeneity, annotationsurvey} or only the tabular modality~\cite{empoweringtabulardatapreparation, shi2025surveysynthetictabular}


\noindent $\bullet$ \textbf{Systematic Taxonomy \textit{vs.} Coarse or Narrow Method Category.}
We propose a unified taxonomy that systematically organizes existing \llm-enhanced methods by underlying techniques, including prompt-based and \llm agent-based frameworks. In contrast, prior surveys either classifies works using coarse, general categories~\cite{llmdata} or limit their focus to specific methods, such as agent-based systems~\cite{dataagent}.


\noindent $\bullet$ \textbf{Paradigm Shift Summary \textit{vs.} Static Description.}
We systematically examine how data preparation has evolved from rule-based systems to LLM agent frameworks, summarizing the corresponding advantages and limitations. In contrast, prior studies~\cite{llmdata} present works individually, offering limited analysis of paradigm shifts and little discussion of the field's evolution.

\noindent $\bullet$ \textbf{Emerging Challenges and Roadmap \textit{vs.} Conventional Perspectives.}
We summarize challenges in the LLM era, including inference costs, hallucinations, and cross-modal consistency, and outline a forward-looking roadmap. This distinguishes our work from existing surveys that focus primarily on typical issues (e.g., scalability) or offer generic conclusions, providing guidance for the next-generation data preparation.

Moreover, we have the following observations on the evolution of methodology across data preparation tasks.

\noindent $\bullet$ \textbf{Shift Toward Cost-Efficient Hybrid Methods.} 
Recent work moves beyond exclusive reliance on LLM inference and instead adopts hybrid approaches. Among them, LLMs either generate executable preparation programs or transfer their reasoning capabilities to smaller language models (SLMs), thereby reducing execution cost and improving scalability.

\noindent $\bullet$ \textbf{Reduced Emphasis on Task-Specific Fine-Tuning.}
The focus has shifted away from maintaining heavily fine-tuned, task-specific LLMs toward methods that optimize other aspects, such as the input construction. Techniques such as retrieval augmentation and structured serialization are used to adapt general-purpose models to new tasks, enabling greater flexibility and lower maintenance overhead.

\noindent $\bullet$ \textbf{Limited Attempts of Agentic Implementations.}
Although agent-based orchestration supports more autonomous data preparation workflows, relatively few systems have been fully studied and implemented in practice. This gap indicates that reliable and robust agentic deployment remains to be explored.

\noindent $\bullet$ \textbf{Task-Specific Methodology Difference.}
Data cleaning employs a hybrid LLM-ML approach for accurate error detection and repair; data integration emphasizes multi-model collaboration to scale matching and alignment; and data enrichment integrates retrieval-augmented and hybrid prompting techniques to enhance the semantic understanding of data and metadata.

\noindent $\bullet$ \textbf{Cross-Modal Generalization with Unified Representations.}
Recent methods increasingly support multiple data modalities within a single architecture. By using shared semantic representations, these methods process tables, text, and other data uniformly, reducing the reliance on modality-specific feature engineering.

\endgroup

\section{Data Preparation: Definition and Scope}




In this section, we provide a clear definition of three fundamental data preparation tasks, including \emph{Data Cleaning} to remove errors and inconsistencies from raw data, \emph{Data Integration} to combine and harmonize data from multiple sources, and \emph{Data Enrichment} to identify patterns, relationships, and knowledge that support informed decisions.



\bfit{Data Cleaning} aims to convert corrupted or low-quality data within a dataset into a trustworthy form suitable for downstream tasks (e.g., statistical analysis). It involves tasks such as fixing typographical errors, resolving formatting inconsistencies, and handling violations of data dependencies. Recent \llm-enhanced studies primarily focus on three critical tasks including data standardization, data error detection and correction of data errors, and data imputation.


\noindent $\bullet$ \emph{\underline{(1) Data Standardization}}~\cite{datastandardization1, datastandardization2} aims to transform heterogeneous, inconsistent, or non-conforming data into a unified representation that satisfies predefined consistency requirements. 
Formally, given a dataset $\mathcal{D}$ and consistency criteria $\mathcal{C}$, it applies or learns a standardization function $f_\mathrm{std}$ such that the output dataset $\mathcal{D}_\mathrm{std}=f_\mathrm{std}(\mathcal{D}, \mathcal{C})$ satisfies $\mathcal{C}$. 
Typical tasks include format normalization (e.g., converting dates from ``7th April 2021'' to ``20210407''), case normalization (e.g., ``SCHOOL'' to ``school''), and symbol or delimiter cleanup (e.g., removing redundant separators ``1000 .'' to obtain ``1000''). 
\llm{}-enhanced methods leverage context-aware prompting and reasoning-driven code synthesis to produce automated, semantically consistent transformations, reducing reliance on manual pattern definition and improving generalization across heterogeneous data formats.



\noindent $\bullet$ \emph{\underline{(2) Data Error Processing}}~\cite{tool2,dataerrorprocessing2, dataerrorprocessing1} refers to the two-stage process of detecting erroneous values and subsequently repairing them to restore data reliability. 
Formally, given a dataset \(\mathcal{D}\) and a set of error types \(\mathcal{K}\), an detection function \(f_{\text{id}}(\mathcal{D}, \mathcal{K})\) identifies an error set \(\mathcal{D}_{\text{err}}\), after which a repair function \(f_{\text{fix}}\) produces a refined dataset \(\mathcal{D}_\mathrm{fix} = f_{\text{fix}}(\mathcal{D}, \mathcal{D}_{\text{err}})\) such that \(f_{\text{id}}(\mathcal{D_\text{fix}}, \mathcal{K}) = \emptyset\).
Typical tasks include identifying data irregularities (e.g., constraint violations) and performing data corrections (e.g., resolving encoding errors) to uphold data correctness.
\llm{}-enhanced techniques employ hybrid LLM–ML architectures and executable code generation to deliver accurate, scalable error identification and correction, thereby lowering dependence on hand-crafted rules and boosting adaptability across varied, noisy datasets.



\noindent $\bullet$ \emph{\underline{(3) Data Imputation}}~\cite{dataimputation1,dataimputation2,dataimputation3} refers to the task of detecting missing data entries and estimating plausible values for them, with the goal of restoring a dataset’s structural completeness and logical coherence.
More formally, given a dataset $\mathcal{D}$ containing missing entries, the objective is to learn or apply an imputation function $f_{\mathrm{imp}}$ that yields a completed dataset $\mathcal{D}_\mathrm{imp} = f_{\mathrm{imp}}(\mathcal{D})$, in which all previously missing entries are filled with inferred, plausible values.
Typical tasks include predicting absent columns based on correlated attributes (e.g., deducing a missing city from a phone area code) or exploiting auxiliary sources (e.g., inferring missing product attributes using relevant tuples from a data lake).
\llm{}-enhanced approaches use semantic reasoning and external knowledge to generate accurate, context-aware replacements, lessening dependence on fully observed training data and enhancing generalization across heterogeneous datasets.


{\bfit{Data Integration} aims to align elements across diverse datasets so that they can be accessed and analyzed in a unified, consistent manner. Instead of exhaustively enumerating all integration task, this survey focuses on entity matching and schema matching, as these are key steps in real-world data integration workflows and have received the most attention in recent LLM-based research.}

\noindent $\bullet$ \emph{\underline{(1) Entity Matching}}~\cite{entitymatching1, entitymatching2} refers to the task of deciding whether two records correspond to the same real-world entity, facilitating data alignment within a single dataset or across multiple datasets.
More formally, given two collections $R_1$ and $R_2$ and a record pair $(r_1, r_2)$ with $r_1 \in R_1$ and $r_2 \in R_2$, the objective is to estimate and assign a score to the likelihood that the two records describe the same entity.
Typical subtasks include mapping product listings across different e-commerce sites (e.g., associating the same item on Amazon and eBay) and detecting duplicate customer entries.
\llm{}-enhanced entity matching leverages structured prompting and collaboration among multiple models to deliver robust and interpretable matching, reducing dependence on task-specific training and enhancing generalization across diverse schemas.

\noindent $\bullet$ \emph{\underline{(2) Schema Matching}}~\cite{schemamatching1,schemamatching2,schemamatching3} aims to identify semantic correspondences between columns or tables across heterogeneous schemas, thereby supporting integrated data access and analysis.
Formally, given a source schema \(\mathcal{S}_s\) and a target schema \(\mathcal{S}_t\), each represented as a collection of tables with their respective column sets, the goal is to learn a mapping function \(f_{\mathrm{sm}}\) that maps every source column $\mathcal{A}_s$ to a semantically equivalent target column $\mathcal{A}_t$ (or to \(\emptyset\) if no suitable counterpart exists).
Common subtasks involve matching columns whose names with synonymous meanings (e.g., linking \texttt{price} in one table with \texttt{cost} in another) and detecting correspondences between tables (e.g., aligning \texttt{CustomerInfo} with \texttt{ClientDetails}).
\llm{}-enhanced schema matching leverages prompt-based reasoning, retrieval-augmented information, and multi-agent coordination to handle semantic ambiguity and structural variation, thereby lowering reliance on hand-crafted rules and improving alignment quality across heterogeneous domains.

\bfit{{Data Enrichment}} focuses on augmenting datasets by adding semantic labels and descriptive metadata, or by discovering complementary datasets that increase their value for downstream tasks (e.g., data analysis). It involves subtasks such as classifying column types and producing dataset-level descriptions. This survey concentrates on data annotation and data profiling, which represent the predominant enrichment operations in existing LLM-enhanced studies.




\noindent $\bullet$ \emph{\underline{(1) Data Annotation}}~\cite{dataannotation1,dataannotation2} aims to attach semantic or structural labels to elements in raw data so that they can be understood and utilized by downstream applications. 
Formally, given a dataset $\mathcal{D}$, the objective is to define a labeling function $f_{\mathrm{ann}}$ that maps each data element to one or more labels in $\mathcal{L}$, such as its semantic role or data type.
Typical subtasks include {semantic column-type annotation} (e.g., identifying a column as \texttt{CustomerID} or \texttt{birthDate}), {table-class detection} (e.g., determining that a table is an \texttt{Enterprise Sales Record}), {and cell entity annotation (e.g., linking the cell \texttt{Apple} to the entity \texttt{Apple\_Inc})}.
\llm{}-enhanced annotation {leverages instruction-based prompting, retrieval-augmented context, and fine-tuning} to deliver precise, scalable, and domain-sensitive labeling, substantially decreasing manual workload and reducing manual effort and mitigating hallucination compared to traditional task-specific models.



\noindent $\bullet$ \emph{\underline{(2) Data Profiling}}~\cite{dataprofiling1, dataprofiling2} refers to the task of systematically analyzing a dataset to derive its structural, statistical, and semantic properties, {as well as identifying associations with relevant datasets,} thereby producing rich metadata that facilitates data comprehension and quality evaluation.
Formally, for a dataset $\mathcal{D}$, a profiling function $f_{\mathrm{pro}}$ generates a metadata collection $\mathcal{M}=\{m_1,\ldots,m_k\}$, where each metadata element $m_i$ encodes characteristics such as distributional statistics, structural regularities, semantic categories, or {connections to semantically related datasets}.
Common subtasks include {semantic metadata generation} (e.g., summarizing the contents of tables and assigning domain-aware descriptions to columns) and {structural relationship extraction} (e.g., clustering related columns and inferring hierarchical dependencies).
\llm{}-enhanced profiling combines prompt-based analysis, retrieval-augmented contextualization, and layered semantic reasoning to yield accurate, interpretable metadata that improves data exploration, enables quality assurance, and offers a reliable foundation for downstream applications.

Unlike data preparation pipelines designed specifically for training, fine-tuning, or directly prompting \llms themselves~\cite{llmdata}, this survey focuses on \llm-enhanced data preparation methods that aim to refine the quality, consistency, and semantic coherence of data used in downstream analytical and machine-learning applications, as summarized in Table~\ref{tab:table-check}.



\section{\llm \space for Data Cleaning}
\label{section:cleaning}

Traditional data cleaning methods rely on rigid rules and constraints (e.g., ZIP code validation), which demand substantial manual effort and domain expertise (e.g., schema knowledge in financial data)~\cite{AutoDCWorkflow, GIDCL}.
Moreover, they often require task-specific training, which limits their generalization across different scenarios~\cite {Evaporate}.
Recent studies show that \llms can address these limitations by reducing manual and programming effort (e.g., offering natural language interfaces), and supporting the seamless integration of domain knowledge for the following tasks. 





\noindent \textbf{Data Standardization.}
Data standardization refers to transforming heterogeneous or non-uniform values into a unified format, enabling dependable analysis and efficient downstream processing.
Existing \llm-enhanced standardization techniques can be classified into three main categories.

\begin{figure}[!t]
  \centering
  \includegraphics[width=\linewidth]{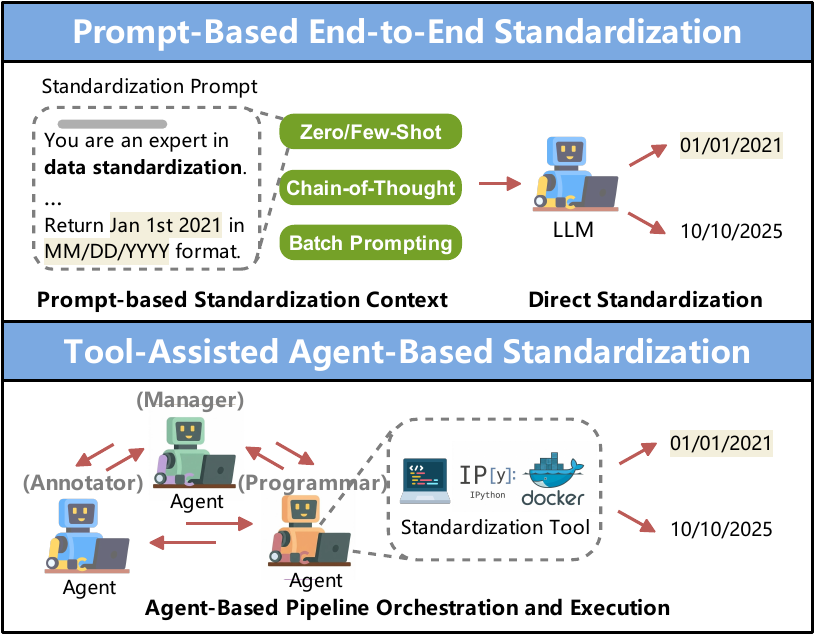}
  \caption{Example of \llm-Enhanced Data Standardization.}
  \label{fig:data_standardization}
  \vspace{-.5cm}
\end{figure}

\noindent \bfit{\ding{182} Prompt-Based End-to-End Standardization.}
As shown in Figure~\ref{fig:data_standardization}, this method uses structured prompts that specify detailed standardization rules (e.g., normalization criteria) or provide stepwise reasoning instructions, guiding \llms to generate data outputs in a standardized format.

\noindent $\bullet$ \underline{\emph{Instruction-Guided Standardization Prompting.}}
{This category relies on manually crafted prompts, together with in-context or labeled standardization examples, to guide \llms in performing data standardization across diverse tasks.}
For instance, LLM-GDO~\cite{LLMGDO} employs user-specified prompts with parameterized templates to encode data standardization rules as textual instructions (e.g., ``convert dates into YYYYMMDD'') and to substitute user-defined functions (e.g., executable formatting code implementations).




\noindent $\bullet$ \underline{\emph{Reasoning-Enhanced Batch Standardization Prompting.}}
This category leverages step-wise reasoning and batch-wise processing prompting to enhance both the standardization robustness and efficiency. 
For instance, LLM-Preprocessor~\cite{LLMPreprocessor} proposes a unified prompting framework that tackles hallucinations, domain shifts, and computational costs through: (1) zero-shot Chain-of-Thought prompting, which elicits step-by-step reasoning to first verify the correct target column and then to guide \llms in producing the standardized output; and (2) batch-wise prompting, which feeds multiple items into a single prompt so they can be processed simultaneously.

\noindent \bfit{\ding{183} Automatic Code-Synthesis Standardization.}
This approach standardizes data by instructing \llms to generate executable code that performs the standardization. 
The generated code is then executed to ensure uniform data handling and improve efficiency.
For instance, Evaporate~\cite{Evaporate} prompts \llms to produce code that derives structured representations from semi-structured documents; results from multiple candidate functions are then combined to boost accuracy while preserving low computational overhead.

\noindent \bfit{\ding{184} Tool-Assisted Agent-Based Standardization.}
As shown in Figure~\ref{fig:data_standardization}, this approach overcomes the challenges of complex prompt design by employing \llm agents to coordinate and execute standardization pipelines.
For instance, CleanAgent~\cite{CleanAgent} maps specific standardization operations with domain-specific APIs, and relies on agents to execute a standardization pipeline, which involves generating API calls (e.g., \texttt{clean\_date(df, "Admission Date", "MM/DD/YYYY")}) and executing them iteratively.
Similarly, AutoDCWorkflow~\cite{AutoDCWorkflow} leverages \llm agents to assemble pipelines and carry out stepwise reasoning to locate relevant columns, evaluate data quality, and apply appropriate operations (e.g., \texttt{upper()} and \texttt{trim()}), while leveraging tools such as OpenRefine~\cite{OpenRefine} for execution and feedback.


\begin{takeawaybox}
\noindent \textbf{Discussion.} 
\bfit{(1) Prompt-Based Standardization for Heterogeneous Modalities.}
This paradigm leverages structured instructions and in-context examples to flexibly convert diverse inputs into a unified format, enabling rapid, training-free deployment~\cite{Evaporate}. Nonetheless, its dependence on direct \llm inference leads to high token consumption and constrains scalability for large-scale or frequently repeated tasks.
\bfit{(2) Code-Based Standardization for Scalable Execution.}
This paradigm enhances efficiency by using reusable transformation functions with fixed execution cost, making it well-suited for processing large datasets~\cite{Evaporate, LLMPreprocessor}. However, it is vulnerable to errors because \llms may produce faulty code, requiring the aggregation of multiple candidate functions to maintain reliability.
\bfit{(3) Agentic-Based Standardization for Automated Pipelines.}
This paradigm constructs automated pipelines by translating natural-language specifications into executable workflows, thereby increasing usability and transparency~\cite{LLMGDO, CleanAgent, AutoDCWorkflow}. However, coordinating numerous tools and APIs introduces additional maintenance overhead and can increase latency relative to direct prompt-based approaches.
\end{takeawaybox}

\noindent \textbf{Data Error Processing.} 
Given a data item, data error processing typically involves two stages: detecting errors and then correcting them.
Common error types include typographical mistakes (typos), anomalous numeric values, and violations of data dependencies.
Existing approaches to error processing can generally be grouped into four major categories.




\noindent \bfit{\ding{182} Prompt-Based End-to-End Error Processing.}
This approach relies on structured prompts that describe explicit error detection and correction instructions, organize processing steps into iterative workflows, or incorporate illustrative examples and reasoning guidance, to instruct \llms to identify and repair data errors directly.


\noindent $\bullet$ \underline{\emph{Instruction-Based Processing Prompting.}}
This category pairs explicit prompting instructions with serialized tabular rows to guide \llms in performing error detection and correction. For instance, Cocoon-Cleaner~\cite{Cocoon} uses batch-style prompting by serializing sampled values from each column (e.g., 1{,}000 entries per column) and grouping them by their corresponding subject column. It allows \llms to iteratively identify and fix issues such as typos and inconsistent formats, with minimal supervision (e.g., five labeled tuples).



\noindent $\bullet$ \underline{\emph{Workflow-Based Iterative Processing Prompting.}}
This category encompasses iterative, multi-step processing workflows (e.g., the detect–verify–repair loop), in which \llm repeatedly executes, evaluates, and refines processing operations.
For instance, LLMErrorBench~\cite{LLMErrorBench} guides \llms through an iterative sequence of dataset examination, targeted correction (e.g., value substitution), and automated quality evaluation, using prompts enriched with contextual cues such as error locations.
To address newly introduced errors and the dependence on rigid, predefined rules in sequential pipelines, IterClean~\cite{IterClean} introduces an integrated prompting framework in which \llms simultaneously serve as error detector, self-verifier, and data repairer within a continuous feedback loop.

\noindent $\bullet$ \underline{\emph{Example- and Reasoning-Enhanced Processing Prompting.}}
This category incorporates few-shot examples and explicit reasoning steps into error-handling pipelines.
For instance, $\text{Multi-News}^{+}$~\cite{MultiNews} employs Chain-of-Thought prompting in conjunction with majority voting and self-consistency verification, thereby mimicking human decision-making to enhance both the accuracy and interpretability of noisy document classification.
To alleviate the need for manually crafting intricate parsing rules for semi-structured data errors, LLM-SSDC~\cite{SSDC} recasts the problem as a text correction task, using a one-shot prompt that includes general instructions and a single input-output example. This allows \llms to automatically fix structural misplacements (e.g., relocating paragraph indices from a \texttt{<content>} tag to a \texttt{<num>} tag).

\noindent \bfit{\ding{183} Function-Synthesis-Oriented Error Processing.} To address the scalability  of manually crafting rules, this approach leverages \llms to synthesize executable processing functions that explicitly encode table semantics and data dependencies. For instance, LLMClean~\cite{LLMClean} instructs \llms to derive a collection of ontological functional dependencies (OFDs) from the dataset schema, the data, and a domain ontology, which together define validation rules within a context model.
Each OFD represents a concrete rule, such as \textit{ZipCode}~$\rightarrow$~\textit{City} in a postal ontology. These OFDs are subsequently used to detect errors (e.g., inconsistent values) and to steer iterative data repair via integrated tools such as Baran~\cite{Baran}.

\begin{figure}[!t]
  \centering
  \includegraphics[width=\linewidth]{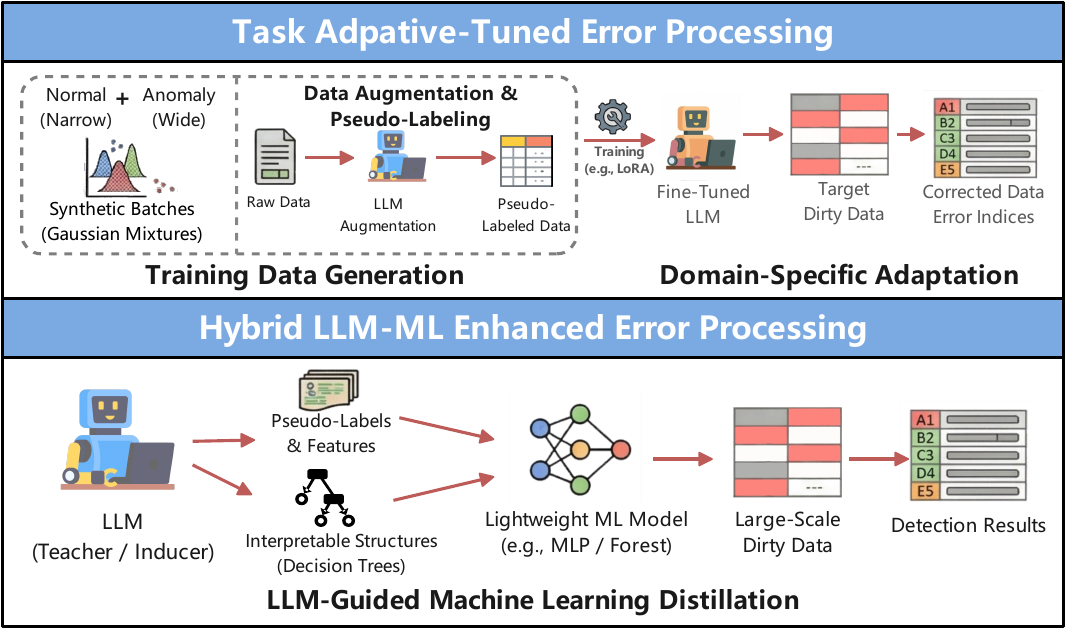}
  \caption{Example of \llm-Enhanced Data Error Processing.}
  \label{fig:data_error_processing}
\end{figure}

\noindent \bfit{\ding{184} Task-Adaptive Fine-Tuned Error Processing.}
As shown in Figure~\ref{fig:data_error_processing}, this method fine-tunes \llms to learn dataset-specific error patterns that are hard to capture via prompting alone, leveraging synthetic noise or contextual augmentation to enhance both error detection and correction performance.


\noindent $\bullet$ \underline{\emph{Synthetic Noise-Augmented Fine-Tuning.}}
This category fine-tunes \llms using synthetic datasets augmented with different noises, such as Gaussian or multinomial, to learn error detection.
For instance, LLM-TabAD~\cite{LLM-TabAD} adapts base \llms (e.g., \texttt{Llama 2}~\cite{Llama2}) for error detection by constructing synthetic datasets where each example is a small batch of rows together with the indices of the abnormal rows.
Continuous columns in the rows are drawn from a mixture of a narrow Gaussian (normal values) and a wide Gaussian (anomalous extremes), while categorical columns are sampled from two multinomial distributions with different probability patterns.
Each batch is then serialized into a natural-language description, and the \llm is fine-tuned to predict the anomaly row indices.


\begin{sloppypar}
\noindent $\bullet$ \underline{\emph{LLM-Based Context Augmentation Fine-Tuning.}}
In this category, \llms are fine-tuned using prompts that are enriched with additional contextual information, such as serialized neighboring cells and retrieved similar examples. As an illustration, GIDCL~\cite{GIDCL} constructs fine-tuning data by combining labeled tuples with pseudo-labeled tuples produced via \llm-based augmentation. Each training instance is represented as a context-enriched prompt that includes: (1) an instruction (e.g., ``Correct the ProviderID to a valid numeric format''), (2) a serialized erroneous cell along with its row and column context (e.g., ``\texttt{\textless COL\textgreater ProviderID\textless VAL\textgreater 1x1303...}''), (3) in-context learning examples (e.g., ``bxrmxngham $\rightarrow$ birmingham''), and (4) retrieval-augmented examples drawn from the same cluster (e.g., clean tuples obtained via $k$-means).
\end{sloppypar}

\noindent \bfit{\ding{185} Hybrid LLM-ML Enhanced Error Processing.}
As shown in Figure~\ref{fig:data_error_processing}, this approach integrates \llms with machine learning models to strike a balance between accuracy and computational efficiency in handling errors. In practical deployments, \llms are either employed to create labeled datasets that train ML models, or to derive structural representations that guide ML-based error processing.

\noindent $\bullet$ \underline{\emph{LLM-Labeled ML Processing Training.}}
In this category, \llm is employed as a data labeler to create pseudo-labels and synthetic examples of correctly identified errors, which are then used to train a lightweight ML model that serves as an efficient detector.
As an illustrative instance, ZeroED~\cite{ZeroED} uses \llms to annotate features and subsequently trains a lightweight ML classifier (e.g., an MLP) for end-to-end error detection.
The training dataset is obtained via a zero-shot pipeline: representative values are first chosen through clustering, then labeled by the \llm, and these labels are propagated to nearby values. The dataset is further enriched with \llm-generated synthetic corruptions {(e.g., substituting valid ages with impossible values such as 999)} to better capture rare error patterns.

\noindent $\bullet$ \underline{\emph{LLM-Induced Structure for ML Processing.}}
In this category, \llm is employed as a logical blueprint to construct interpretable error-detection programs, which are later run and combined by machine-learning models.
As an illustration, to enhance both explainability and robustness in data processing, ForestED~\cite{ForestED} restructures the processing pipeline by leveraging the \llm to produce transparent decision structures (e.g., trees whose nodes apply rule-based format or range checks, along with relational nodes that encode cross-column dependencies), while downstream ML models execute and aggregate these structures to generate the final predictions.

\begin{takeawaybox}
\textbf{Discussion.}
\bfit{(1) Prompt-Based Processing for End-to-End Workflows.}
This approach reframes error processing as a generative modeling problem through data serialization~\cite{SSDC, MultiNews}, and couples decomposed pipelines with iterative verification loops to ensure robust reasoning~\cite{Cocoon, IterClean, LLMErrorBench}. However, direct \llm inference remains constrained by token limits when operating on large tables, and iterative self-correction cycles can compound hallucinations or introduce new errors.
\bfit{(2) Function-Synthesis Processing for Automatic Rule Discovery.}
This paradigm leverages \llms to autonomously identify hidden dependencies and synthesize explicit, executable cleaning routines directly from raw data~\cite{LLMClean}.
However, deriving strict validation rules from already corrupted inputs risks overfitting to noise, causing the \llm to synthesize invalid rules that effectively encode errors as valid rules.
\bfit{(3) Task-Adaptive Error Processing for Domain Specificity.}
This strategy addresses the text–table modality discrepancy by fine-tuning \llms on synthetic noise or context-enriched datasets to capture complex, dataset-specific error patterns~\cite{LLM-TabAD, GIDCL}. Nonetheless, it requires a significant ``cold start'' investment to curate or generate sufficiently high-quality training data. 
\bfit{(4) Hybrid \llm{}-ML Error Processing for Scalable Detection.}
This approach uses \llms to produce pseudo-labels~\cite{ZeroED} or to derive interpretable decision structures~\cite{ForestED} that guide lightweight, scalable ML classifiers. 
However, the ultimate detection performance is tightly constrained by both the fidelity of the initial \llm-generated labels and the capabilities of the induced models.
\end{takeawaybox}

\noindent \textbf{Data Imputation.}
For a data record that contains missing entries (e.g., {null values}), data imputation aims to estimate these unknown values using the surrounding contextual information.
Existing \llm-enhanced approaches can be grouped into three main categories.

\begin{figure}[!t]
  \centering
  \includegraphics[width=\linewidth]{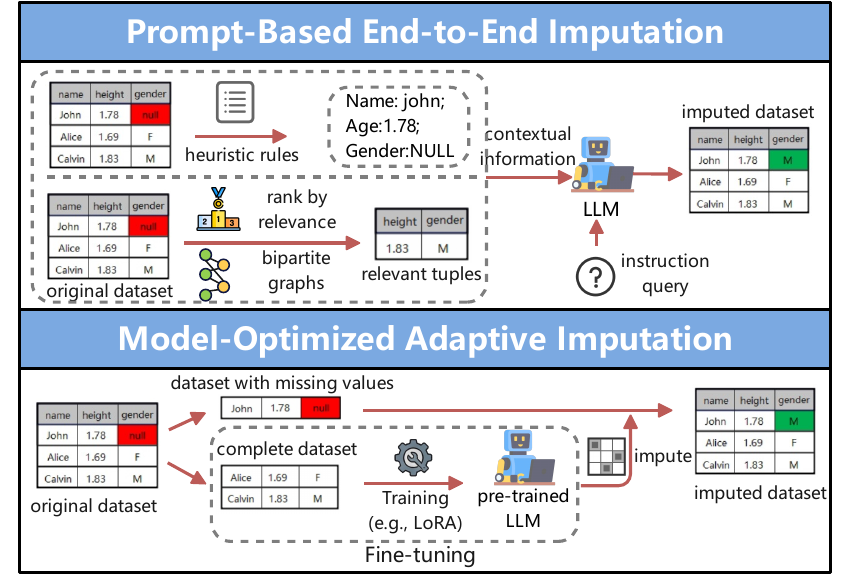}
  \caption{Example of \llm-Enhanced Data Imputation.}
  \label{fig:data_imputation}
\end{figure}

\noindent \bfit{\ding{182} Prompt-Based End-to-End Imputation.} 
As shown in Figure~\ref{fig:data_imputation}, this approach uses structured prompts to direct \llms to fill in missing values in a single step. Existing methods either arranges imputation prompts via heuristic formatting schemes or selectively augments prompts with relevant context.

\noindent $\bullet$ \underline{Heuristic-Structured Imputation Prompting.}
This category organizes imputation prompts using heuristic rules that aim to optimize the formatting of instructions for missing value imputation.
For instance, CRILM~\cite{CRILM} employs rule-based prompt design by converting feature names into natural language phrases (e.g., turning alcohol into ``alcohol content''), retaining the observed values (e.g., 12.47), and adding domain-specific context (e.g., wine). These components are then combined into explicit natural language statements such as ``The alcohol content in the wine is 12.47''.
The resulting descriptions are supplied as prompts to \llms, along with detailed instructions for producing descriptions for the missing values.

\noindent $\bullet$ \underline{Selective Imputation Context Prompting.}
This category focuses on including only the most relevant information in the imputation context, thereby reducing redundancy and token usage.
For instance, LLM-PromptImp~\cite{LLM-PromptImp} refines the context by choosing the columns that are most relevant to the target missing attribute, where relevance is determined using correlation metrics (e.g., Pearson correlation, Cramer's V, and $\eta$ correlation) tailored to different data types.
LDI~\cite{LDI2025} narrows the imputation context by first detecting columns that exhibit explicit dependency relationships with the target column, and then selecting a small number of representative tuples whose values are among the top-$k$ most similar to the incomplete tuple, measured by the normalized length of the longest common substring across these dependent columns.
\llm-Forest~\cite{LLMForest} enables selective construction of the imputation context by converting tabular data into hierarchically merged bipartite information graphs and then retrieving neighboring nodes that are both correlated and diverse for tuples containing missing entries.

\noindent \bfit{\ding{183} Context-Retrieval Guided Imputation.}
This approach enables \llms to handle previously unseen, domain-specific, or private datasets by dynamically enriching the input with supplemental context retrieved from external sources.
{For instance, RetClean~\cite{RetClean} builds an index over a data lake using both syntactic and semantic retrieval, selects a pool of candidate tuples, reranks them with a learned ranking model, and then presents the dirty tuple together with the top-$k$ retrieved tuples to \llms for imputation.}
Similarly, LakeFill~\cite{LakeFill} adopts a two-stage retriever–reranker architecture: an initial vector-based retriever assembles a broad candidate set from the data lake, followed by a reranker that filters this down to a small set of highly relevant tuples that form the imputation context.






\noindent \bfit{\ding{184} Model-Optimized Adaptive Imputation.}
As shown in Figure~\ref{fig:data_imputation}, this approach improves imputation quality by adjusting either the \llm's training procedure or its architecture to better capture complex relationships in mixed-type tabular data.

\noindent $\bullet$ \underline{Adaptive Model Fine-Tuning Optimization.}
This category improves imputation by fine-tuning \llms on task-specific datasets through parameter-efficient methods.
For example, LLM-REC~\cite{LLMREC} adopts a data-partitioned fine-tuning framework that divides the dataset into complete and incomplete portions. It then leverages the complete portion to partially fine-tune the \llm using LoRA, thereby enabling the model to impute missing values based on the observed data.

\noindent $\bullet$ \underline{Module-Augmented Architecture Optimization.}
This class of methods incorporates dedicated modules into \llms to model structural or feature-level dependencies that standard \llms may overlook.
For instance, UnIMP~\cite{highorderim} augments the \llm with two lightweight components that capture interactions among numerical, categorical, and textual cells: (1) a high-order message-passing module that aggregates both local and global relational information, and (2) an attention-based fusion module that merges these features with prompt embeddings prior to decoding the final imputed values. 
Building on UnIMP, Quantum-UnIMP~\cite{QuantumUnIMP2025} adds a quantum feature-encoding module that maps mixed-type inputs into classical vectors used to parameterize an Instantaneous Quantum Polynomial (IQP) circuit. The resulting quantum embeddings serve as the initial node representations in the UnIMP hypergraph.

\begin{takeawaybox}
\bfit{Discussion.} \bfit{(1) Prompt-Based Imputation for Balanced Efficiency.}
This line of work leverages structured prompts and targeted context removal to reduce token consumption while mitigating class imbalance~\cite{LLM-PromptImp}.
However, aggressive pruning can omit subtle cross-column relationships that are crucial for inferring missing values in complex, high-dimensional tables.
\bfit{(2) Retrieval-Guided Imputation for Verifiable Privacy.}
This paradigm relies on RAG to ground imputation in external data lakes, enabling explicit source attribution and secure, on-premise deployment~\cite{RetClean}.
However, its performance is tightly constrained by the coverage and fidelity of relevant tuples in the data lake, and retrieval noise can further impair imputation accuracy.
\bfit{(3) Model-Optimized Imputation for Structural Complexity.}
This strategy incorporates tailored architectural components or incremental training schemes to model global and local dependencies in heterogeneous, mixed-type datasets~\cite{highorderim}. 
Nonetheless, these specialized components introduce additional architectural complexity and higher computational costs compared to standard, general-purpose \llms.
\end{takeawaybox}


\section{\llm for Data Integration}
\label{section:integration}

Traditional integration methods often struggle with semantic ambiguities and inconsistencies, especially in complex settings where domain-specific knowledge is unavailable~\cite{KGRAG4SM}.
Moreover, pretrained language models generally demand substantial task-specific training data and often suffer from performance degradation when dealing with out-of-distribution entities~\cite{MatchGPT}.
By contrast, recent work has demonstrated that \llms exhibit strong semantic understanding, allowing them to detect relationships across datasets and integrate domain knowledge, thereby achieving robust generalization across a wide range of integration tasks.





\noindent \textbf{Entity Matching.}
Entity matching aims to decide whether a pair of data records corresponds to the same real-world entity.
Existing \llm-enhanced approaches can be broadly grouped into three main categories.






\noindent \bfit{\ding{182} Prompt-Based End-to-End Matching.} 
This approach relies on structured prompts to guide \llms in performing entity matching directly. 
Existing methods either include explicit guidance via detailed instructions and in-context examples or organize candidate tuples into batches to enhance efficiency.

\noindent $\bullet$ \underline{\emph{Guidance-Driven In-Context Matching Prompting}.} 
This category enhances entity matching through carefully structured in-context guidance, including strategically selected demonstrations, expert-defined logical rules, and multi-step prompting pipelines. 
For example, MatchGPT~\cite{MatchGPT} prepares guidance by selecting in-context demonstrations via various strategies (e.g., similarity-based vs. manual) and automatically generating textual matching rules from handwritten examples.
ChatEL~\cite{ChatEL} further follows the guidance of a multi-step pipeline to first retrieve candidates, then generate task-oriented auxiliary descriptions, and finally perform instruction-guided multiple-choice selection to identify matches.
To mitigate hallucination and reliance on the \llm's internal knowledge, KcMF~\cite{KcMF} incorporates expert-designed pseudo-code of if-then-else logic enriched with external domain knowledge, and employs an ensemble voting mechanism to aggregate multi-source outputs.



\noindent $\bullet$ \underline{\emph{Batch-Clustering Matching Prompting}.}
This category enhances matching efficiency by packing multiple entities or entity pairs into a single prompt, allowing \llms to jointly reason about them.
For instance, BATCHER~\cite{BATCHER} groups multiple entity pairs into one prompt via a greedy, cover-based selection strategy that clusters pairs exhibiting similar matching semantics (e.g., relying on the same matching rules or patterns).
Similarly, LLM-CER~\cite{LLM-CER} employs a list-wise prompting approach that processes a batch of tuples at once, using in-context examples to cluster related entities in a single pass and thereby lowering the cost associated with sequential pairwise matching.

\begin{figure}[!t]
  \centering
  \includegraphics[width=\linewidth]{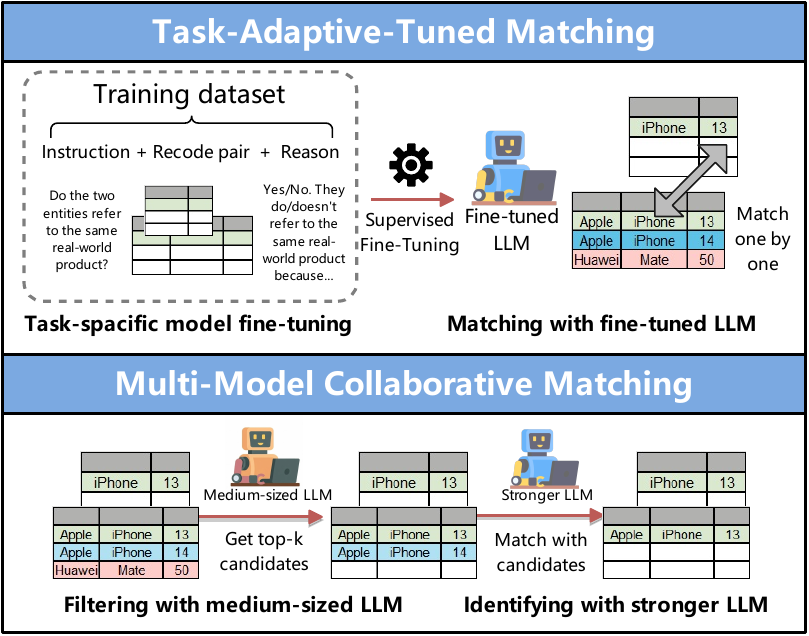}
  \caption{Example of \llm-Enhanced Entity Matching.}
  \label{fig:entity_matching}
\end{figure}

\noindent \bfit{\ding{183} Task-Adaptive-Tuned Matching.}
As shown in Figure~\ref{fig:entity_matching}, this approach fine-tunes \llms for entity matching using task-specific supervision, either by distilling reasoning traces from stronger models or by improving training data quality to enhance matching adaptability and generalization.

\noindent $\bullet$ \underline{\emph{Reasoning-Distilled Matching Tuning}.} 
This category fine-tunes local small \llms using Chain-of-Thought traces distilled from larger models.
For example, Jellyfish~\cite{Jellyfish} performs parameter-efficient instruction tuning on small models (ranging 7B-13B) using reasoning traces (derived from CoT prompting over serialized data) distilled from a larger mixture-of-experts \llm (e.g., Mixtral-8x7B) to improve reasoning consistency and task transferability.

\noindent $\bullet$ \underline{\emph{Data-Centric Matching Tuning}.} 
This category optimizes the fine-tuning process by improving the quality of training data via enriched information.
For example, FTEM-LLM~\cite{FTEM-LLM} adds clear explanations to the training data that describe why two items are the same or different (e.g., comparing specific columns). It also cleans the data by removing mislabeled examples and generating hard negatives via embedding-space neighbor selection. 
Similarly, LLM-CDEM~\cite{LLM-CDEM} demonstrates that data-centric strategies {(e.g., Anymatch~\cite{Anymatch} uses an AutoML-based strategy to identify and add hard examples to the training set, and uses attribute-level augmentation to increase the training set's granularity)}, which focus on improving training data quality, significantly outperform model-centric approaches in achieving robust cross-domain generalization.

\noindent \bfit{\ding{184} Multi-Model Collaborative Matching.}
As shown in Figure~\ref{fig:entity_matching}, this approach enhances entity matching by coordinating multiple models to exploit their complementary strengths.
For instance, COMEM~\cite{COMEM} proposes \llm collaboration in a combined local and global matching strategy, where a medium-sized \llm (3B-11B) ranks top-$k$ candidates via bubble sort to mitigate position bias and context-length dependency, and a stronger \llm (e.g., GPT-4o) refines these candidates by modeling inter-tuple interactions to ensure globally consistent and accurate matching.
{To effectively resolve long-tail entity ambiguity and maintain computational efficiency, LLMaEL~\cite{LLMaEL}~leverages \llms as context augmenters to generate entity descriptions as additional input for small entity matching models. 
The augmented context is integrated via concatenation, fine-tuning, or ensemble methods to guide small entity matching models to produce accurate results.}

\begin{takeawaybox}
\textbf{Discussion.} \bfit{(1) Prompt-Based Matching for End-to-End Resolution.} 
This approach utilizes structured guidance (e.g., logical rules, multi-step pipelines)~\cite{MatchGPT, ChatEL, KcMF} and batching strategies~\cite{BATCHER, LLM-CER} to perform matching directly, facilitating explainable decisions and improved efficiency.
However, reliance on the \llms' internal knowledge makes it sensitive to input phrasing and incurs significant token costs for large-scale candidate lists.
\bfit{(2) Task-Adaptive Matching for Robust Adaptation.} This approach bridges the gap between security and generalization by fine-tuning local models~\cite{Jellyfish} or prioritizing data-centric training strategies to handle unseen schemas~\cite{FTEM-LLM, LLM-CDEM}. 
However, it faces a significant ``cold start'' challenge, requiring high-quality, diverse training data to prevent overfitting or performance regression on out-of-distribution domains. 
\bfit{(3) Multi-Model Collaborative Matching for Scalable Consistency.} 
This approach leverages lightweight rankers for preliminary blocking~\cite{COMEM} or context augmentation~\cite{LLMaEL} to address position bias and global consistency violations. 
However, the pipeline's overall accuracy is strictly bounded by the recall of the preliminary blocking stage, as early filtering errors cannot be recovered by the \llm.
\end{takeawaybox}

\noindent \textbf{Schema Matching.} 
The objective of schema matching is to identify correspondences between elements across different database schemas (e.g., matching column names such as ``employee ID'' and ``staff number'').
Existing \llm-enhanced approaches can be divided into five categories.

\begin{figure}[!t]
  \centering
  \includegraphics[width=\linewidth]{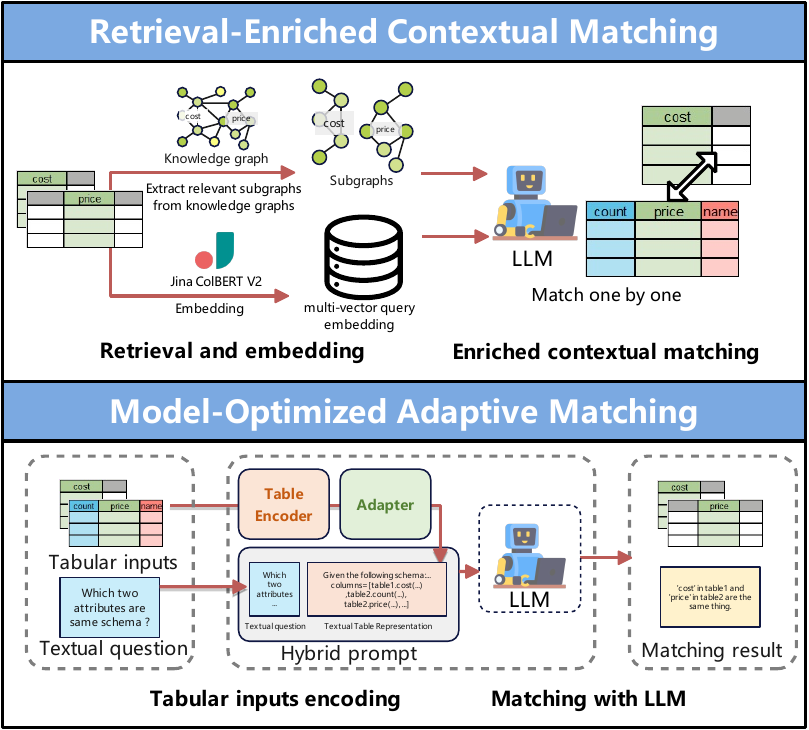}
  \caption{Example of \llm-Enhanced Schema Matching.}
  \label{fig:schema_matching}
  \vspace{-.8cm}
\end{figure}

\noindent \bfit{\ding{182} Prompt-Based End-to-End Matching.}
This approach uses structured prompts to enable \llms to perform schema matching without explicit code implementations. For example, LLMSchemaBench~\cite{LLMSchemaBench} designs prompts for different tasks across varying contexts and adopts prompting patterns such as persona specification (e.g., instructing \llms to act as a schema matcher), match-criteria definition, Chain-of-Thought reasoning instructions, and structured output formats.
GLaVLLM~\cite{GLaVLLM} further optimizes matching prompts by three strategies: (1) it improves output consistency by applying symmetric transformations to the input schemas and aggregating multiple outputs; (2) it increases matching expressiveness through structured prompting and rule decomposition, supporting complex matching patterns such as ``Global-and-Local-as-View'', where multiple source relations jointly define multiple target relations; and (3) it reduces token usage by filtering tasks based on data types and grouping similar tasks before prompting \llms.

\noindent \bfit{\ding{183} Retrieval-Enriched Contextual Matching.}
As shown in Figure~\ref{fig:schema_matching}, this approach improves schema matching by augmenting \llm inputs with context obtained from external retrieval components.
For instance, Matchmaker~\cite{Matchmaker} integrates pretrained retrieval models (such as ColBERTv2~\cite{Colbert}) with \llms by encoding columns at the token level for vector-based semantic retrieval, and then using an \llm to score and rank the retrieved candidates.
KG-RAG4SM~\cite{KGRAG4SM} extends this idea by employing multiple retrieval strategies, including vector-based, graph traversal, and query-driven search—to extract relevant subgraphs from knowledge graphs, which are then ranked and injected into \llm prompts to provide richer context for matching.

\noindent \bfit{\ding{184} Model-Optimized Adaptive Matching.}
As shown in Figure~\ref{fig:schema_matching}, this approach enhances matching effectiveness through modality-aware fine-tuning, complemented by specialized module designs. 
For example, TableLlama~\cite{tablellama} applies instruction tuning over a wide range of table-centric tasks, allowing the model to learn alignment strategies and column semantics implicitly, without changing its core architecture.
Building on this, TableGPT2~\cite{tablegpt2} adopts an architecture-augmented optimization scheme by incorporating a two-dimensional table encoder that generates permutation-invariant representations, thereby enhancing the stability and accuracy of cross-table column alignment and candidate match ranking.



\noindent \bfit{\ding{185} Multi-Model Collaborative Matching.}
This approach improves schema matching by coordinating multiple models with complementary capabilities. For example, Magneto~\cite{Magneto} adopts a retrieve-and-rerank framework in which small pre-trained language models first produce candidate match rankings for each input column, and \llms subsequently refine these candidates through reranking to achieve higher matching accuracy and efficiency.

\noindent \bfit{\ding{186} Agent-Guided Orchestration-Based Matching.} 
In this paradigm, \llm agents are used to manage and coordinate the entire schema matching pipeline.
Existing methods either designate distinct agents to handle and carry out specific matching subtasks or depend on agent-based planning mechanisms to orchestrate a set of predefined tools.

\noindent $\bullet$ \underline{\emph{Role-Based Matching Orchestration}.}
In this category, the workflow is partitioned into specialized agents, each responsible for different operations.
For instance, Agent-OM~\cite{AgentOM} uses two \llm agents (a Retrieval Agent and a Matching Agent) to coordinate the matching process, breaking tasks down via Chain-of-Thought prompting, calling specialized tools (such as syntactic, lexical, and semantic retrievers and matchers), and relying on a hybrid memory architecture (relational + vector database) for storage and retrieval.

\noindent $\bullet$ \underline{\emph{Tool-Planning Matching Orchestration}.}
This category uses \llm agents to coordinate predefined tools through dynamic planning to solve complex matching problems.
For example, Harmonia~\cite{Harmonia} employs \llm agents to orchestrate and integrate a set of predefined data integration tools (i.e., modular algorithms tailored to specific matching subtasks, such as \texttt{top\_matches} for retrieving the top-$k$ most suitable candidates), and complements them with on-demand code generation when the available tools are inadequate. At the same time, it incorporates mechanisms such as ReAct~\cite{yao2023react} for joint reasoning and action planning, interactive user feedback for correcting errors, and declarative pipeline specifications to guarantee reproducibility.

\begin{takeawaybox}
\textbf{Discussion.} \bfit{(1) Prompt-Based Matching for Stable Alignment.}
This paradigm employs one-to-many comparisons and symmetric transformations to promote consistency and reduce sensitivity to inputs constrained by privacy~\cite{LLMSchemaBench, GLaVLLM}.
However, when it relies exclusively on metadata, the model is unable to interpret semantically opaque column names, and its exhaustive verification strategy leads to prohibitive token consumption for large schemas.
\bfit{(2) Retrieval-Enriched Matching for Hallucination Resistance.}
This approach grounds the alignment in verifiable semantic subgraphs by retrieving contextual information from external knowledge graphs~\cite{KGRAG4SM}.
However, its performance can be constrained by the domain coverage of the external knowledge source and the added retrieval overhead (e.g., graph traversal). 
\bfit{(3) Model-Optimized Matching for Structural Semantics.}
This approach integrates specialized architectural components (e.g., table encoders) or task-oriented fine-tuning to encode table-specific alignment regularities~\cite{tablellama, tablegpt2}.
However, it relocates the bottleneck to training data acquisition, demanding high-quality or large-scale datasets to achieve robust generalization across heterogeneous domains.
\bfit{(4) Multi-Model Matching for Cost-Efficient Scale.}
This paradigm relies on \llms to generate training instances for lightweight scorers, forming a scalable filter-then-rank pipeline~\cite{Magneto}.
However, the ultimate matching quality is tightly constrained by the fidelity of the synthetic training data and the loss of reasoning capability transferred to the smaller model.
\bfit{(5) Agent-Guided Matching for Autonomous Workflows.}
This approach leverages chain-based reasoning and self-refinement strategies to coordinate complex, multi-stage alignment procedures~\cite{AgentOM}.
HoweverNonetheless, the complex orchestration of tools and iterative reasoning cycles can introduce additional latency and maintenance overhead compared with static \llms.
\end{takeawaybox}


\section{\llm for Data Enrichment}
\label{section:enrichment}

Existing data enrichment techniques suffer from two main drawbacks.
First, they  limited interactions between queries and tables~\cite{Birdie}.
Second, many such methods depend strongly on large labeled corpora, are brittle under distribution shifts, and do not generalize well to rare or highly specialized domains~\cite{ArcheType, LLMCTA}.
Recent studies have shown that \llms can mitigate these issues by producing high-quality metadata, enhancing the contextual information of datasets, and enabling natural language interfaces for performing enrichment tasks.

\noindent \textbf{Data Annotation.}
Data annotation is the process of assigning semantic or structural labels to data instances, such as identifying column types (e.g., \texttt{Manufacturer} or \texttt{birthDate} in the DBPedia ontology).
Recent \llm-enhanced methods typically can be divided into five main categories.




\noindent \bfit{\ding{182} Prompt-Based End-to-End Annotation.}
This approach utilizes carefully crafted prompts to guide \llms in performing diverse annotation tasks. It involves methods that supply explicit annotation guidelines and contextual information, while also leveraging reasoning and iterative self-refinement to improve annotation accuracy.


\noindent $\bullet$ \underline{\emph{Instruction-Guided Annotation Prompting.}}
This category uses structured prompts with explicit instructions to guide \llms in performing data annotation tasks. For example, CHORUS~\cite{CHORUS} designs prompts that combine correct annotation demonstrations, serialized data samples, metadata, domain knowledge, and output formatting guidance. 
Similarly, EAGLE~\cite{EAGLE} employs task-specific prompts to selectively label critical or uncertain samples (identified via prediction disagreement), combining zero-shot \llm annotation with active learning to enhance generalization in low-data settings.
ArcheType~\cite{ArcheType} adopts a column-at-once serialization strategy that includes only representative column samples for zero-shot column type annotation. To handle abbreviated column names, Columbo~\cite{Columbo} defines prompt instructions over three modules: (1) a summarizer module generates concise group and table summaries from context to provide annotation guidance, (2) a generator module expands tokenized column names into meaningful phrases, and (3) a reviser module evaluates and refines the consistency of these expanded phrases.

\noindent $\bullet$ \underline{\emph{Reasoning-Enhanced Iterative Annotation Prompting.}}
This category enhances annotation quality by using structured prompts that guide models through step-by-step reasoning and iterative self-assessment to produce more accurate labels.
For example, Goby~\cite{Goby} applies tree-structured serialization and Chain-of-Thought prompting for enterprise column type annotation.
AutoLabel~\cite{AutoLabel} performs automated text annotation on representative samples (selected via DBSCAN~\cite{DBSCAN} clustering and stratified sampling) using domain-optimized CoT reasoning templates that decompose complex labeling tasks into stepwise instructions (e.g., ``First classify entity types, then assess confidence levels''), while a human feedback loop iteratively validates low-confidence outputs. 
Anno-lexical~\cite{Anno-lexical} further adopts a majority voting mechanism that aggregates annotations from multiple open-source \llms to enhance annotation robustness and reduce bias.
LLMCTA~\cite{LLMCTA} produces and iteratively improves label definitions using prompt-driven methods, such as self-refinement (progressively enhancing definitions by learning from errors) and self-correction (a two-stage process involving a separate reviewer model). LLMLog~\cite{LLMLog} tackles ambiguity in log template generation via multi-round annotation, leveraging self-evaluation metrics like prediction confidence to identify uncertain or representative logs, and repeatedly updating in-context examples to refine prompt instructions and boost annotation accuracy.

\noindent \bfit{\ding{183} RAG-Assisted Contextual Annotation.}
This approach enriches \llm prompts to enhance annotation by retrieving relevant context, either from semantically similar examples or from external knowledge graphs.

\noindent $\bullet$ \underline{\emph{Semantic-Based Annotation Example Retrieval.}}
This category enhances annotation accuracy by retrieving semantically relevant examples to enrich the prompt context.
For instance, LLMAnno~\cite{LLMAnno} addresses the inefficiency of manually selecting examples for large-scale named entity recognition (e.g., annotating 10,000 resumes) by retrieving the most relevant training examples and constructing context-enriched prompts for \llms. Experiments show that retrieval based on appropriate embeddings (e.g., \texttt{text-embedding-3-large}~\cite{openai2024embeddings}) outperforms zero-shot and in-context learning across multiple \llms (7B-70B parameters) and datasets.

\noindent $\bullet$ \underline{\emph{Graph-Based Annotation Knowledge Retrieval.}}
This category enhances annotation by retrieving relevant entity triples from external knowledge graphs to enrich the prompt context. For example, RACOON~\cite{RACOON} extracts entity-related knowledge (e.g., labels and triples) from a knowledge graph, converting it into concise contextual representations, and incorporating it into prompts to enhance semantic type annotation accuracy.

\noindent \bfit{\ding{184} Fine-Tuned Augmented Annotation.}
This approach improves annotation in specialized domains by fine-tuning \llms on task-specific datasets. For example, PACTA~\cite{PACTA} combines low-rank adaptation with prompt augmentation, decomposing prompts into reusable patterns and training across diverse contexts to reduce prompt sensitivity in column type annotation. OpenLLMAnno~\cite{OpenLLMAnno} demonstrates that fine-tuned local \llms (e.g., \texttt{Llama 2}, \texttt{FLAN-T5}) outperform proprietary models like GPT-3.5 in specialized text annotation tasks, achieving substantial accuracy gains even with a small number of labeled samples (e.g., 12.4\% improvement with 100 samples for \texttt{FLAN-T5-XL}).

\begin{figure}[!t]
  \centering
  \includegraphics[width=\linewidth]{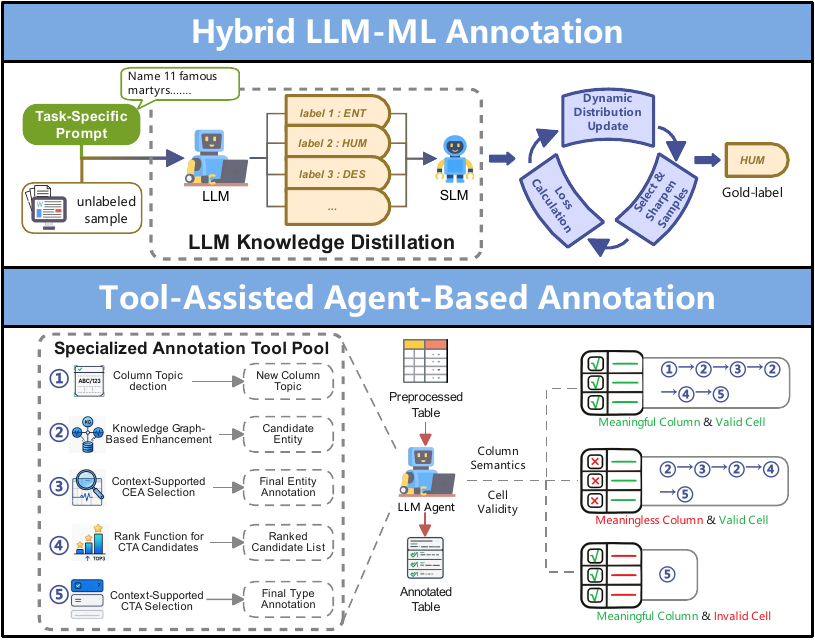}
  \caption{Example of \llm-Enhanced Data Annotation.}
  \label{fig:data_annotation}
  \vspace{-.8cm}
\end{figure}

\noindent \bfit{\ding{185} Hybrid LLM-ML Annotation.}
As shown in Figure~\ref{fig:data_annotation}, this approach combines \llms with ML models to improve annotation accuracy and robustness through knowledge distillation and collaborative orchestration. For instance, CanDist~\cite{CanDist} employs a distillation-based framework where \llms uses task-specific prompts to generate multiple candidate annotations, and SLMs (e.g., RoBERTa-Base) then distill and filter them.
A distribution refinement mechanism updates the SLM's distribution, gradually correcting false positives and improving robustness to noisy data. 
AutoAnnotator~\cite{AutoAnnotator} uses two-layer collaboration: (1) \llms act as meta-controllers, selecting suitable SLMs, generating annotation, and verifying hard samples, while (2) SLMs perform bulk annotation, produce high-confidence labels via majority voting, and iteratively fine-tune on \llm-verified hard samples to enhance generalization.

\noindent \bfit{\ding{186} Tool-Assisted Agent-Based Annotation.}
As shown in Figure~\ref{fig:data_annotation}, this approach uses \llm agents augmented with specialized tools to handle complex annotation tasks. For example, STA Agent~\cite{STA-Agent} leverages a ReAct-based \llm agent for semantic table annotation, combining preprocessing (e.g., spelling correction, abbreviation expansion) with tools for column topic detection, knowledge graph enrichment, and context-aware selection, while reducing redundant outputs via Levenshtein distance. TESSA~\cite{TESSA} employs a multi-agent system for cross-domain time series annotation, integrating general and domain-specific agents with a multi-modal feature extraction toolbox for intra- and inter-variable analysis and a reviewer module to ensure consistent and accurate annotations.

\begin{takeawaybox}
\textbf{Discussion.} \bfit{(1) Prompt-Based Annotation for Complex Reasoning.}
This approach uses structured prompts to capture iterative feedback~\cite{LLMCTA, CHORUS} or multi-step reasoning~\cite{Goby, Columbo}, progressively refining annotation guidelines to clarify ambiguous schemas. However, the reliance on lengthy, complex instructions and repeated interactions might lead to high token consumption and latency.
\bfit{(2) Retrieval-Enriched Annotation for Factual Accuracy.}
This approach fetches context from external knowledge bases to ground annotations in verifiable information, enabling more reliable handling of specialized domains where the model's internal knowledge may be obsolete~\cite{RACOON, LLMAnno}.
However, its accuracy is tightly constrained by the reliability of the external resources and by noise from irrelevant or low-quality retrieved content.
\bfit{(3) Fine-Tuned Annotation for Domain Specificity.}
This approach adapts open-source models for specific domains (e.g., law, politics) via parameter-efficient fine-tuning, reaching high accuracy with lower deployment costs~\cite{OpenLLMAnno}.
However, this merely shifts the primary bottleneck to data acquisition, since it requires extensive, high-quality instruction data to avoid overfitting.
\bfit{(4) Hybrid LLM-ML Annotation for Scalable Deployment.}
This approach trains lightweight ML models on weighted label distributions generated by \llms, ensuring cost-effective inference~\cite{CanDist}. However, the ML model's performance is fundamentally limited by the \llms's upper bound, and the distillation step often results in a loss of the reasoning depth required for subtle edge cases.
\bfit{(5) Agent-Based Annotation for Tool-Assisted Tasks.}
This approach uses autonomous agents that call external tools (e.g., search engines) for resolving hard-to-label entities~\cite{STA-Agent}.
However, the sequential use of multiple tools might introduce significant delays, making it impractical for real-time or high-volume annotation.
\end{takeawaybox}

\noindent \textbf{Data Profiling.}
{Data profiling involves characterizing a given dataset by generating additional information (e.g., dataset descriptions, schema summaries, or hierarchical organization) or associating relevant datasets that enrich its structural and semantic understanding.}
Recent \llm-enhanced methods can be classified into two categories.



\begin{figure}[!t]
  \centering
  \includegraphics[width=\linewidth]{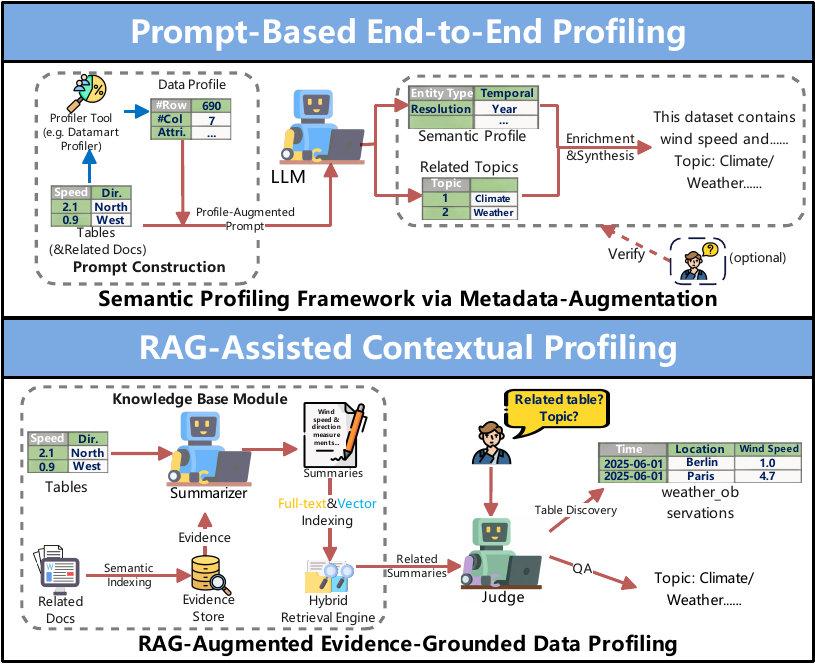}
  \caption{Example of \llm-Enhanced Data Profiling.}
  \label{fig:data_profiling}
  \vspace{-.5cm}
\end{figure}

\noindent \bfit{\ding{182} Prompt-Based End-to-End Profiling.}
As shown in Figure~\ref{fig:data_profiling}, this approach uses carefully designed prompts to guide \llms in profiling datasets, combining explicit instructions or constraints with few-shot examples and reasoning to handle complex, heterogeneous, and structured data effectively.

\noindent $\bullet$ \underline{\emph{Instruction and Constraint-Based Profiling Prompting.}}
This category guides dataset profiling by incorporating explicit instructions or usage constraints in prompts to cover various aspects of the data.
For example, AutoDDG~\cite{AutoDDG} instructs \llms to generate both user-oriented and search-optimized descriptions based on dataset content and intended usage. LEDD~\cite{LEDD} employs prompts with task-specific instructions for data lake profiling, including summarizing clusters into hierarchical categories and refining natural language queries for semantic search. DynoClass~\cite{DynoClass} specifies instructions in the prompt to synthesize detailed table descriptions from sampled rows and existing documentation, integrating them into a coherent global hierarchy. LLM-HTS~\cite{LLM-HTS} instructs \llms to infer open-set semantic types for tables and columns, which are then used to build hierarchical semantic trees via embedding-based clustering. Cocoon-Profiler~\cite{10.1145/3665939.3665957} describes instructions at three levels: (1) table-level prompts constrain summarization using initial rows and documentation, (2) schema-level prompts guide hierarchical column grouping in JSON format, and (3) column-level prompts generate descriptions based on example rows and global context.
HyperJoin~\cite{HyperJoin} instructs \llms to create semantically equivalent column name variants using table context and naming conventions, producing structured JSON outputs to construct inter-table hyperedges.
OCTOPUS~\cite{Octopus} specifies strict constraints in the prompts to output only column names separated by specific delimiters and a SQL sketch, enabling lightweight entity-aware profiling.

\noindent $\bullet$ \underline{\emph{Example and Reasoning-Enhanced Profiling Prompting.}}
This category combines few-shot example prompts with Chain-of-Thought (CoT) reasoning to support structured profiling of complex and heterogeneous data. For instance, LLMCodeProfiling~\cite{LLMCodeProfiling} uses a two-stage, prompt-based framework for cross-language code profiling. In the syntactic abstraction stage, few-shot CoT prompts demonstrate how abstract syntax tree (AST) nodes from different languages can be converted into a unified tabular representation, guiding the \llm to infer deterministic mappings that align language-specific constructs to a common schema. In the semantic assignment stage, instructional classification prompts direct the \llm to assign imported packages to functional categories (e.g., labeling \texttt{scikit-learn} as ``machine learning'').

\noindent \bfit{\ding{183} RAG-Assisted Contextual Profiling.}
As shown in Figure~\ref{fig:data_profiling}, this approach combines multiple retrieval techniques with \llm reasoning to improve profiling accuracy and consistency, especially when metadata is sparse or incomplete.
For example, LLMDap~\cite{LLMDap} employs vector search to gather relevant textual evidence, including scientific articles, documentation, and metadata fragments, to generate semantically consistent dataset-level profiles (e.g., dataset descriptions, variable definitions, and structured metadata). Pneuma~\cite{Pneuma} integrates hybrid retrieval methods, such as full-text and vector search, to identify relevant tables from databases or data lakes, using \llms to generate semantic column descriptions and to refine and rerank the retrieved results.


\begin{takeawaybox}
\textbf{Discussion.} \bfit{(1) Prompt-Based Profiling for Descriptive Summarization.}
This approach integrates structural and statistical metadata into prompts to generate faithful dataset descriptions, overcoming context window limits~\cite{AutoDDG}.
However, relying solely on summary statistics is a lossy compression, potentially causing the model to miss fine-grained semantic anomalies hidden in the raw data.
\bfit{(2) Iterative Profiling for Hierarchical Structure.}
This approach utilizes \llm-driven clustering and summarization to build hierarchical views of data lakes, enabling semantic search across disparate tables~\cite{LEDD}.
However, the iterative abstraction process risks accumulating information loss, resulting in vague or generic descriptions at higher levels of the hierarchy.
\bfit{(3) Hybrid Profiling for Quality Assurance.}
This approach augments statistical profiling with \llm-driven reasoning and human verification to identify complex structural anomalies and disguised missing values~\cite{Cocoon}.
However, the reliance on human-in-the-loop intervention creates a scalability bottleneck, making it unsuitable for fully automated, real-time data pipelines.
\bfit{(4) Retrieval-Enriched Profiling for Contextual Grounding.}
This approach retrieves external context (e.g., similar tables or text) to ground the generation of evidence-based schema descriptions~\cite{Pneuma, LLMDap}. However, the final profiling accuracy is strictly bounded by the relevance of the retrieved corpus, where noisy or outdated external context can induce hallucinations.
\end{takeawaybox}

\section{Evaluation}
\label{sec:evaluation}

\subsection{\ds Datasets}
\label{subsed:datasets}

\begin{table*}[!htbp]
\centering
\vspace{-.4cm}
\caption{Summary of Representative \ds Datasets.
}
\label{tab:datasets_summary}
\resizebox{\linewidth}{!}{
\begin{tabular}{|c|c|c|c|c|c|c|}
\hline
\textbf{Category} &
  \textbf{Dataset} &
  \textbf{Task} &
  \textbf{Modality} &
  \textbf{Granularity} &
  \textbf{Data Volume (Unit)} &
  \textbf{Evaluation Metric} \\ \hline
\multirow{14}{*}{\begin{tabular}[c]{@{}c@{}}Data\\ Cleaning\end{tabular}} &
  Chicago Food Inspection~\cite{chicagoopendata, AutoDCWorkflow} &
  DS &
  Tabular &
  column &
  298,345 (rows) &
  Precision, Recall, F1-score \\ \cline{2-7} 
 &
  Paycheck Protection Program~\cite{ppp, AutoDCWorkflow} &
  DS &
  Tabular &
  column &
  661,218 (rows) &
  Precision, Recall, F1-score \\ \cline{2-7} 
 &
  Enron Emails~\cite{enron, Evaporate} &
  DS &
  Text &
  tuple &
  517,401 (emails) &
  F1-score \\ \cline{2-7} 
 &
  SWDE Movie / University / NBA~\cite{swde, Evaporate} &
  DS &
  Other &
  document / page &
  20,000 / 16,705 / 4,405 (pages) &
  F1-score \\ \cline{2-7} 
 &
  Hospital~\cite{dataerrorprocessing2, LLMPreprocessor, Cocoon, AutoDCWorkflow, GIDCL, ForestED, ZeroED, IterClean} &
  DS, DEP &
  Tabular &
  cell / tuple / column &
  1,000 (rows) &
  Precision, Recall, F1-score \\ \cline{2-7} 
 &
  Flights~\cite{flights, CleanAgent, AutoDCWorkflow, Cocoon, GIDCL, ForestED, ZeroED, IterClean} &
  DS, DEP &
  Tabular &
  cell / tuple / column &
  2,377 (rows) &
  Precision, Recall, F1-score, Matching Rate \\ \cline{2-7} 
 &
  Beers~\cite{beers, Cocoon, GIDCL, ForestED, ZeroED, IterClean} &
  DEP &
  Tabular &
  cell / tuple / column &
  2,410 (rows) &
  Precision, Recall, F1-score \\ \cline{2-7} 
 &
  Meat Consumption~\cite{meat_consumption, LLMErrorBench} &
  DEP &
  Tabular &
  table &
  12,140 (rows) &
  F1-score \\ \cline{2-7} 
 &
  Hotel Booking~\cite{hotel_booking, LLMErrorBench} &
  DEP &
  Tabular &
  table &
  119,390 (rows) &
  F1-score \\ \cline{2-7} 
 &
  Adult Income~\cite{adult_income, LLMPreprocessor, LLM-PromptImp} &
  DEP, DI &
  Tabular &
  cell / tuple / tuple &
  48,842 (rows) &
  Precision, Recall, F1-score, Accuracy, ROC-AUC \\ \cline{2-7} 
 &
  Travel datasets~\cite{travel_customer_churn, LLM-PromptImp} &
  DI &
  Tabular &
  tuple &
  954 (rows) &
  Precision, Recall, F1-score, Accuracy, ROC-AUC \\ \cline{2-7} 
 &
  Buy~\cite{abt-buy, IPM, LLMPreprocessor, highorderim} &
  DI &
  Tabular &
  cell &
  651 (rows) &
  Accuracy, ROUGE-1, Cos-Sim \\ \cline{2-7} 
 &
  Restaurant~\cite{riddleDatasets2003, IPM, LLMPreprocessor, highorderim} &
  DI &
  Tabular &
  cell &
  864 (rows) &
  Accuracy, ROUGE-1, Cos-Sim \\ \cline{2-7} 
 &
  Walmart~\cite{magellandata, IPM, highorderim} &
  DI &
  Tabular &
  cell &
  4,654 (rows) &
  ROUGE-1, Cos-Sim \\ \hline
\multirow{10}{*}{\begin{tabular}[c]{@{}c@{}}Data\\ Integration\end{tabular}} &
  abt-buy~\cite{abt-buy, Jellyfish, LLM-CDEM, COMEM, MatchGPT, BATCHER} &
  EM &
  Tabular &
  tuple pair &
  1,097 (pairs) &
  F1-score \\ \cline{2-7} 
 &
  Amazon-Google~\cite{abt-buy, LLMPreprocessor, Jellyfish, LLM-CDEM, COMEM, MatchGPT, BATCHER} &
  EM &
  Tabular &
  tuple pair &
  1300 (pairs) &
  F1-score \\ \cline{2-7} 
 &
  Walmart-Amazon~\cite{magellandata, LLMPreprocessor, Jellyfish, LLM-CDEM, COMEM, MatchGPT, BATCHER} &
  EM &
  Tabular &
  tuple pair &
  1154 (pairs) &
  F1-score \\ \cline{2-7} 
 &
  DBLP-Scholar / ACM~\cite{abt-buy, LLMPreprocessor, Jellyfish, LLM-CDEM, COMEM, MatchGPT, BATCHER} &
  EM &
  Tabular &
  tuple pair &
  5347 / 2224 (pairs) &
  F1-score \\ \cline{2-7} 
 &
  WDC Products~\cite{WDC_products, MatchGPT, LLM-CDEM} &
  EM &
  Tabular &
  tuple pair &
  40,500 (pairs) &
  F1-score \\ \cline{2-7} 
 &
  OMOP~\cite{OMOP, KGRAG4SM} &
  SM &
  Tabular &
  attribute-pair &
  37 (tables), 394 (attributes) &
  Precision, Recall, F1-score, Accuracy \\ \cline{2-7} 
 &
  Synthea~\cite{Synthea, LLMPreprocessor, KGRAG4SM, KcMF, Matchmaker, GLaVLLM} &
  SM &
  Tabular &
  attribute-pair &
  12 (tables), 111 (attributes) &
  Precision, Recall, F1-score, Accuracy \\ \cline{2-7} 
 &
  MIMIC~\cite{mimic, KGRAG4SM, KcMF, Matchmaker, GLaVLLM} &
  SM &
  Tabular &
  attribute-pair &
  25 (tables), 240 (attributes) &
  Precision, Recall, F1-score, Accuracy \\ \cline{2-7} 
 &
  GDC-SM~\cite{GDC-SM, Magneto} &
  SM &
  Tabular &
  attribute-pair &
  20 (tables) &
  MRR, Recall@GT \\ \cline{2-7} 
 &
  ChEMBL-SM~\cite{ChEMBL, Magneto} &
  SM &
  Tabular &
  attribute pair &
  8 (datasets) &
  MRR, Recall@GT \\ \hline
\multirow{10}{*}{\begin{tabular}[c]{@{}c@{}}Data\\ Enrichment\end{tabular}} &
  NQ-Tables~\cite{NQ-Tables} / OpenWikiTable~\cite{Open-WikiTable, Birdie} &
  DA &
  Tabular &
  table &
  952 / 6,602 (tables) &
  P@k \\ \cline{2-7} 
 &
  DBpedia Ontology Dataset~\cite{DBpedia2015, CanDist} &
  DA &
  Text &
  document &
  70,000 (documents) &
  Accuracy, 1-$\alpha$, F1-score \\ \cline{2-7} 
 &
  AGNews~\cite{AGNews, CanDist} &
  DA &
  Text &
  document &
  7,600 (documents) &
  Accuracy, 1-$\alpha$, F1-score \\ \cline{2-7} 
 &
  CoNLL-2003~\cite{CoNLL, LLMAnno} &
  DA &
  Text &
  document &
  386 (documents) &
  F1-score \\ \cline{2-7} 
 &
  WNUT-17~\cite{WNUT2017, LLMAnno} &
  DA &
  Text &
  document &
  1,287 (documents) &
  F1-score \\ \cline{2-7} 
 &
  ChEMBL-DP~\cite{ChEMBL, Pneuma, Octopus} &
  DP &
  Tabular &
  table &
  78 (tables) &
  Precision, Recall, F1-score, Hit Rate \\ \cline{2-7} 
 &
  Adventure Works~\cite{adventureworks, Pneuma, Octopus} &
  DP &
  Tabular &
  table &
  88 (tables) &
  Precision, Recall, F1-score, Hit Rate \\ \cline{2-7} 
 &
  Public BI Benchmark~\cite{publicbibenchmark, Pneuma, Octopus} &
  DP &
  Tabular &
  table &
  203 (tables) &
  Precision, Recall, F1-score, Hit Rate \\ \cline{2-7} 
 &
  Chicago Open Data~\cite{chicagoopendata, Pneuma, Octopus} &
  DP &
  Tabular &
  table &
  802 (tables) &
  Precision, Recall, F1-score, Hit Rate \\ \cline{2-7} 
 &
  FetaQA~\cite{FeTaQA, Pneuma, Octopus} &
  DP &
  Tabular &
  table &
  10,330 (tables) &
  Precision, Recall, F1-score, Hit Rate \\ \hline
\end{tabular}
}
\\[2pt]
\textit{Abbreviations:} 
DS – Data Standardization; 
DEP – Data Error Processing; 
DI – Data Imputation; 
EM – Entity Matching;  \\
SM – Schema Matching; 
DP – Data Profiling; 
DA – Data Annotation.\\
\end{table*}


{To support a systematic evaluation of \llm-enhanced data preparation, we summarize representative datasets in Table~\ref{tab:datasets_summary}, providing detailed information across multiple dimensions, including category, task, modality, granularity, data volume, and evaluation metrics.
It allows researchers to compare and select benchmarks tailored to their specific use cases.
For instance, we present a \emph{granularity-driven perspective} below that groups benchmarks by their fundamental processing unit (i.e., records, schemas, or entire objects).

\noindent\textbf{(1) Record-Level.} 
This category treats individual \emph{tuples}, \emph{cells}, or \emph{tuple pairs} as the analysis unit. 
It covers most data cleaning, error processing, data imputation, and entity matching tasks, including detecting erroneous values, standardizing attributes, imputing missing cells, and identifying coreference across records. 
Representative \emph{tuple-level} benchmarks include \textit{Adult Income}~\cite{adult_income}, \textit{Hospital}~\cite{dataerrorprocessing2}, \textit{Beers}~\cite{beers}, \textit{Flights}~\cite{flights}, and text-based datasets such as \textit{Enron Emails}~\cite{enron}. 
\emph{Column-level} benchmarks include the \textit{Paycheck Protection Program}~\cite{ppp} and \textit{Chicago Food Inspection}~\cite{chicagoopendata}.
Cell-level benchmarks include \textit{Buy}~\cite{abt-buy}, \textit{Restaurant}~\cite{riddleDatasets2003}, and \textit{Walmart}~\cite{magellandata}. 
Conversely, \emph{tuple-pair} benchmarks, including \textit{abt-buy}~\cite{abt-buy}, \textit{Amazon--Google}~\cite{abt-buy}, \textit{Walmart--Amazon}~\cite{magellandata}, \textit{DBLP--Scholar}~\cite{abt-buy}, \textit{DBLP--ACM}~\cite{abt-buy}, and \textit{WDC Products}~\cite{WDC_products}, focus on pairwise comparisons across heterogeneous sources for record-level alignment.

\noindent\textbf{(2) Schema-Level.} 
This category focuses on \emph{attribute pairs} or \emph{schema elements}, aiming to align columns and conceptual entities across heterogeneous schemas. 
The challenge shifts from validating individual values to matching semantic meanings and structural roles. Benchmarks such as \textit{OMOP}~\cite{OMOP}, \textit{Synthea}~\cite{Synthea}, and \textit{MIMIC}~\cite{mimic} focus on clinical attribute alignment. Moreover, datasets such as \textit{GDC-SM}~\cite{GDC-SM} and \textit{ChEMBL-SM}~\cite{ChEMBL} evaluate cross-source attribute alignment within complex scientific and biomedical schemas.

\noindent\textbf{(3) Object-Level.} 
This category deals with entire \emph{tables} or \emph{documents} as the fundamental processing unit. Unlike record- or schema-level tasks, these benchmarks require reasoning over global structure and broader context. Table-level datasets supporting data profiling and annotation include \textit{Public BI}~\cite{publicbibenchmark}, \textit{Adventure Works}~\cite{adventureworks}, \textit{ChEMBL-DP}~\cite{ChEMBL}, \textit{Chicago Open Data}~\cite{chicagoopendata}, \textit{NQ-Tables}~\cite{NQ-Tables}, and \textit{FetaQA}~\cite{FeTaQA}. \emph{Document-level} benchmarks, such as \textit{AGNews}~\cite{AGNews}, \textit{DBpedia}~\cite{DBpedia2015}, \textit{CoNLL-2003}~\cite{CoNLL}, and \textit{WNUT-17}~\cite{WNUT2017}, require combining evidence across full texts for semantic grounding and annotation.
}

\subsection{\ds Metrics}
\label{subsec:metrics}




{In real deployments, data preparation methods are evaluated across multiple dimensions.
Therefore, we organize evaluation metrics in Table~\ref{tab:datasets_summary} by the aspects they measure, including correctness, robustness, ranking quality, and semantic consistency, rather than only by the tasks.



\noindent \bfit{\ding{182} {Preparation Correctness Assessment}.}
This category evaluates the correctness of preparation methods by measuring how accurately they process target data elements relative to ground-truth references.

\noindent $\bullet$ \underline{\emph{Operation Precision}.}
These metrics quantify the reliability of predictions from preparation methods.
For example, \emph{(1) Accuracy}~\cite{PRFA} measures the proportion of correctly classified elements across relevant and irrelevant elements, commonly used in classification tasks such as error identification in data error processing.
\emph{(2) Precision}~\cite{PRFA} measures the fraction of correctly identified matches or errors among all elements flagged by the method, reflecting output reliability in tasks like entity or schema matching. \emph{(3) F1-score}~\cite{PRFA} extends precision to penalize both incorrect identifications and missed detections within a single measure, making it suitable for applications where both erroneous outputs and overlooked cases are significant.

\noindent $\bullet$ \underline{\emph{Operation Coverage}.}
These metrics reflect whether preparation methods comprehensively address all required elements.
For example, \emph{(1) Recall}~\cite{PRFA} measures the proportion of correctly identified matches or errors among all ground-truth elements, reflecting a method's ability to avoid missed detections in tasks such as entity matching.
\emph{(2) Matching Rate}~\cite{abt-buy} quantifies the proportion of target elements that are successfully aligned to a valid representation, commonly used in tasks such as entity matching.


\noindent \bfit{\ding{183} {Preparation Robustness Assessment}.}
This category evaluates the stability and reliability of preparation methods over diverse datasets.
These metrics measure how consistently a method maintains its effectiveness across varying data distributions and structural complexity.
For example, \emph{(1) ROC}~\cite{FRA} characterizes the trade-off between correctly identifying target elements (e.g., valid matches) and incorrectly flagging non-target elements as the decision threshold varies, providing a global view of method behavior in tasks such as data error processing.
\emph{(2) AUC}~\cite{FRA} summarizes this behavior into a single measure that reflects a method's ability to distinguish relevant from irrelevant elements across all thresholds and is commonly used in tasks such as data error processing.




\noindent \bfit{\ding{184} {Enrichment and Ranking Quality Assessment}.}
This category evaluates the quality of preparation methods by measuring how effectively they retrieve and prioritize relevant information over ground-truth results.

\noindent $\bullet$ \underline{\emph{Retrieval Ranking Quality.}}
These metrics assess the relevance of top-ranked candidates in retrieval-based preparation tasks.
For example, \emph{(1) P@k}~\cite{PRMH} measures the fraction of queries where a correct result is found within the top-$k$ elements, reflecting retrieval utility in data profiling.
\emph{(2) MRR}~\cite{PRMH} measures the average rank position of the first correct result across queries, indicating how quickly relevant elements are placed at the top of the list.

\noindent $\bullet$ \underline{\emph{Enrichment Completeness.}}
These metrics measure how comprehensively preparation methods find all relevant information during data enrichment.
For example, \emph{(1) Recall@GT}~\cite{PRMH} measures the fraction of correctly identified elements among the top-$k$ results, where $k$ is the total number of true elements, assessing coverage in tasks such as entity or schema matching.
\emph{(2) $1-\alpha$}~\cite{he2024candidate} measures the fraction of data elements for which the correct label is present in the set of candidates, evaluating label coverage in tasks such as data annotation.
\emph{(3) Hit Rate}~\cite{PRMH} measures the fraction of search queries that return at least one correct result, evaluating basic retrieval success in tasks such as data annotation.



\noindent \bfit{\ding{185} {Semantic Preservation Assessment}.}
This category evaluates the ability of preparation methods to preserve semantic meaning in the generated outputs.
These metrics measure how consistently a method maintains semantics between its outputs and the reference content.
For example, \emph{(1) ROUGE}~\cite{ROUGE} assesses semantic consistency at the lexical level by measuring $n$-gram overlap between the output and the reference text, commonly used to evaluate whether the outputs retain key terms in tasks such as data standardization.
\emph{(2) Cosine Similarity}~\cite{CS} measures semantic alignment in an embedding space by comparing vector representations of the generated and reference texts with a continuous measure in tasks such as data profiling.

\section{Challenges and Future Directions}
\label{sec:future}

\subsection{Data Cleaning}

\noindent \ding{182} \textbf{Global-Aware and Semantically Flexible Cleaning.}
Most existing prompt-based cleaning methods operate on restrictive local contexts, such as individual rows or small batches~\cite{LLMPreprocessor, Cocoon}.
While retrieval-augmented methods expand this scope by fetching external evidence~\cite{RetClean, LakeFill}, they remain centered on instance-level context and cannot capture dataset-level properties (e.g., uniqueness constraints or aggregate correlations) essential for issues requiring holistic views.
\textit{Future work should explore hybrid systems that integrate \llms with external analysis engines capable of providing global statistics and constraints, enabling joint reasoning over local instances and dataset-level signals while preserving the semantic flexibility.}

\noindent \ding{183} \textbf{Robust and Error-Controlled Cleaning.}
Agent-based data cleaning mimics human-style workflows and can improve cleaning coverage~\cite{CleanAgent, AutoDCWorkflow}, but current systems lack effective safeguards against error accumulation and hallucinated cleaning.
Although recent general-purpose frameworks introduce uncertainty estimation~\cite{LLMUncertainty} and self-correction strategies~\cite{LLMSelfCorrection} to improve agent reliability, these techniques are mostly heuristic and cannot be directly applied to data cleaning tasks that require strict correctness guarantees.
\textit{An important open direction is to design uncertainty-aware agent-based cleaning frameworks that use conservative decision strategies, formal validation mechanisms, and explicit risk control, allowing systems to balance cleaning coverage with measurable error risk and move toward provably robust cleaning pipelines.}

\noindent \ding{185} \textbf{Efficient and Scalable Collaborative Cleaning.}
Prompt-based data cleaning methods struggle to scale to large tables due to context limits~\cite{LLMErrorBench, ZeroED}, while agent-based workflows often incur high computational cost and latency~\cite{CleanAgent}.
Although smaller, locally deployable models and federated learning frameworks enable privacy-preserving cleaning deployments~\cite{LLMFederatedScope}, existing systems lack principled strategies for coordinating models with different capabilities.
\textit{An important future direction is to design hierarchical cleaning frameworks that assign routine cleaning tasks to small local models and reserve \llms for complex reasoning, combined with efficient table partitioning and selective context management to reduce cost and latency without sacrificing cleaning quality.}


\subsection{Data Integration}

\noindent \ding{182} \textbf{Universal and Cross-Domain Integration.}
Recent structure-aware matching methods~\cite{LLMMatcher} and cross-dataset integration studies~\cite{LLM-CDEM} have shown encouraging results, but they generally assume the presence of reasonably informative schemas. In practice, many integration scenarios involve extreme heterogeneity, including unclear or abbreviated attribute names, substantial structural mismatches (e.g., nested data mapped to flat tables), and datasets with little or no usable metadata. These conditions remain difficult for current methods to handle reliably.
\textit{An key future direction is to develop techniques that rely less on schema descriptions and prompts, and instead infer semantic correspondences directly from data instances (e.g., value distributions and co-occurrence patterns), enabling robust integration even when schema information is missing or misleading.}

\noindent \ding{183} \textbf{Universal Integration in Diverse Realistic Datasets.}
Despite recent progress, \llm-enhanced integration methods often require curated examples~\cite{MatchGPT, ChatEL} or domain-specific fine-tuning~\cite{Jellyfish, FTEM-LLM} to achieve high performance.
Although zero-shot cross-domain integration has received increasing attention~\cite{ZeroNER}, it remains limited in realistic integration with varying schema design, value formats, or domain-specific semantics.
Thus, building a single matcher that can reliably transfer integration behaviors across diverse datasets remains a major challenge.
\textit{We should explore research in meta-learning and synthetic data generation to create universal integration models that generalize to new domains without requiring expensive, domain-specific training data.}

\noindent \ding{184} \textbf{Rule-Constrained and Globally Valid Integration.}
Recent in-context clustering methods~\cite{BATCHER} for entity matching can efficiently enforce simple global properties, such as transitivity, during matching~\cite{LLM-CER, KcMF}.
In practice, however, data integration often requires satisfying more complex and domain-specific constraints, including multi-entity relationships, temporal ordering, and business rules.
These constraints are difficult to express and enforce using prompt-based approaches.
\textit{An important future direction is to augment \llm-based integration pipelines with explicit reasoning components, such as constraint solvers and graph-based inference modules, that can be invoked by \llm agents to ensure that integration results respect complex, domain-specific constraints.}

\subsection{Data Enrichment}

\noindent \ding{182} \textbf{Interactive Human-in-the-Loop Enrichment.}
Fully automated data enrichment is often impractical, especially when enrichment decisions are ambiguous or domain dependent~\cite{AutoDDG, LLMDap}. In practice, effective workflows require close collaboration between human experts and \llm-enhanced systems.
However, most existing methods are designed for one-shot automation and provide limited support for interactive refinement, where users can guide decisions, verify results, and correct errors during the enrichment process.
\textit{We need to develop novel interactive frameworks where \llms can explain their reasoning, solicit feedback on ambiguous cases, and incrementally refine enrichment tasks based on human guidance, treating the user as a core component of the system.}

\noindent \ding{183} \textbf{Multi-Aspect and Open-Ended Enrichment.}
Evaluating \llm-enhanced data enrichment remains challenging in two aspects.
First, enrichment often involves multiple aspects, such as annotating column types~\cite{CHORUS, LLMCTA}, expanding textual descriptions~\cite{AutoDDG, LEDD}, which are difficult to assess with a single task-level metric.
Second, many enrichment outputs are free-form text, where quality cannot be judged using simple binary or precision-based measures.
As a result, existing benchmark is largely designed for structured or closed-form tasks and fail to reflect the quality and usefulness of real-world enrichment results.
\textit{A key future direction is to develop standardized enrichment benchmarks that support multi-aspect evaluation and richer assessment criteria, combining automatic metrics with reference-based, model-based, or human-in-the-loop evaluation to better capture enrichment quality, usefulness, and cost in realistic scenarios.}

\noindent \ding{184} \textbf{Faithful and Evidence-Grounded Enrichment.}
Generative data enrichment using \llms can produce fluent but unsupported outputs, such as inferred constraints, textual summaries, or data profiles, particularly when the input data is noisy or incomplete~\cite{Goby, Octopus}. Although retrieval-augmented generation provides useful grounding mechanisms~\cite{LLMAnno, RACOON, Pneuma}, existing approaches are primarily designed for structured tables and do not directly meet the needs of unstructured data enrichment. As a result, enriched content often lacks clear links to the data or knowledge sources that justify it.
\textit{An important future direction is to design faithfulness-aware enrichment methods in which every generated output is explicitly grounded in verifiable evidence, such as supporting data samples, query execution results, or cited external knowledge, so that enriched information is both useful and trustworthy.}

\section{Conclusion}
\label{sec:conclusion}

In this survey, we present a task-centric review of recent advances in \llm-enhanced data preparation, covering data cleaning, data integration, and data enrichment.  We systematically analyze how \llms reshape traditional data preparation workflows by enabling capabilities such as instruction-driven automation, semantic-aware reasoning, cross-domain generalization, and knowledge-augmented processing.
Through a unified taxonomy, we organize representative methods, distill their design principles, and discuss the limitations of existing \llm-enhanced methods.
We also summarize representative datasets and metrics to facilitate comprehensive evaluations of these methods.
Finally, we identify open challenges and outline future research directions.

\clearpage
\newpage

\balance
\scriptsize

    

\bibliographystyle{IEEEtran}

\bibliography{ref/DA}

\end{document}